\documentclass[referee]{raa}            

\usepackage[a4paper=true,dvipdfm=true,pagebackref=true]{hyperref}
\hypersetup{colorlinks = true, linkcolor = green, anchorcolor = red, citecolor = blue, filecolor = red, pagecolor = red, urlcolor = red}
\usepackage{graphicx,times}             
\usepackage{natbib}
\usepackage{amssymb,amsmath}
\bibpunct{(}{)}{;}{a}{}{,}
\usepackage{boxedminipage}

\begin{document}

  \title{Estimation of stellar atmospheric parameters from LAMOST DR8 low-resolution spectra with 20$\leq$SNR$<$30
}

   \volnopage{Vol.0 (20xx) No.0, 000--000}      
   \setcounter{page}{1}          

   \author{Xiangru Li 
      \inst{1}
   \and Zhu Wang
      \inst{2}
   \and Si Zeng
      \inst{2}
    \and Caixiu Liao
      \inst{2}
      \and Bing Du
      \inst{3,4}
      \and X. Kong
      \inst{3,4}
      \and Haining Li
      \inst{3}
}

   \institute{School of Computer Science, South China Normal University, No. 55 West of Yat-sen Avenue, Guangzhou 510631, China, xiangru.li@gmail.com\\
        \and
             School of Mathematical Sciences, South China Normal University, No. 55 West of Yat-sen Avenue, Guangzhou 510631, China\\
        \and
             Key Laboratory of Optical Astronomy, National Astronomical Observatories, Chinese Academy of Sciences, Beijing 100101, China\\
         \and
         University of Chinese Academy of Sciences, Beijing 100049, China
\vs\no
   }

\abstract{
The accuracy of the estimated stellar atmospheric parameter decreases evidently with the decreasing of spectral signal-to-noise ratio (SNR) and there are a huge amount of this kind observations, especially in case of SNR$<$30. Therefore, it is helpful to improve the parameter estimation performance for these spectra and this work studied the ($T_\texttt{eff}, \log~g$, [Fe/H]) estimation problem for LAMOST DR8 low-resolution spectra with 20$\leq$SNR$<$30. We proposed a data-driven method based on machine learning techniques.
Firstly, this scheme detected stellar atmospheric parameter-sensitive features from spectra by the Least Absolute Shrinkage and Selection Operator (LASSO), rejected ineffective data components and irrelevant data.
Secondly, a Multi-layer Perceptron (MLP) method was used to estimate stellar atmospheric parameters from the LASSO features.
Finally, the performance of the LASSO-MLP was evaluated by computing and analyzing the consistency between its estimation and the reference from the APOGEE (Apache Point Observatory Galactic Evolution Experiment) high-resolution spectra. Experiments show that the Mean Absolute Errors (MAE) of $T_\texttt{eff}, \log~g$, [Fe/H] are reduced from the LASP (137.6 K, 0.195 dex, 0.091 dex) to LASSO-MLP (84.32 K, 0.137 dex, 0.063 dex), which indicate evident improvements on stellar atmospheric parameter estimation.
In addition, this work estimated the stellar atmospheric parameters for 1,162,760  low-resolution spectra with 20$\leq$SNR$<$30 from LAMOST DR8 using LASSO-MLP, and released the estimation catalog, learned model, experimental code, trained model, training data and test data for scientific exploration and algorithm study.
}

   \authorrunning{X.R. Li et al.}            
   \titlerunning{Estimation of stellar atmospheric parameters}  

   \maketitle

\section{INTRODUCTION}           
\label{sect:intro}
The stellar atmospheric parameters are important reference for understanding the properties of stars, as well as fundamental information for investigating the formation and evolution of galaxies. Therefore, it is an essential problem to estimate stellar atmospheric parameters from spectra in a large-scale sky survey. At the same time, with the continuous development of large-scale sky surveys, the amount of observed spectra is increasing, especially the amount of spectra observed by the Large Sky Area Multi-Object Fiber Spectroscopic Telescope (LAMOST). LAMOST is a typical spectroscopic telescope, with a wide field of view and the highest spectral acquisition rate in the world, provide abundant observed spectra.

A series of researches have been conducted for estimating stellar atmospheric parameters from spectra of LAMOST (\cite{2017ApJ...836....5H,2017MNRAS.464.3657X,2019ApJS..245...34X,2020ApJS..246....9Z}). However, these studies mainly train models on the spectra with medium and high SNR. For example,     \cite{2017ApJ...836....5H} trained the Cannon on the spectra from LAMOST DR2 with SNR $>$ 100. \cite{2017MNRAS.464.3657X} trained a multiple-linear regression method on the spectra from LAMOST DR2 with SNR $>$ 50.  \cite{2019ApJS..245...34X} trained the DD-Payne on the spectra from LAMOST DR5 with SNR$>$50.
\cite{2020ApJS..246....9Z} trained the SLAM on the spectra from LAMOST DR5 with SNR$>$100. As a result, the performance of learned model on the spectra with low SNR decreases evidently. For example, \cite{2017ApJ...836....5H} showed in Figure 8, the uncertainty of $T_\texttt{eff}$ is greater than 90K, $\log~g$ is greater than 0.17dex, [Fe/H] is greater than 0.11dex in case of SNR$<$30.

\begin{figure}
 \includegraphics[width=7cm]{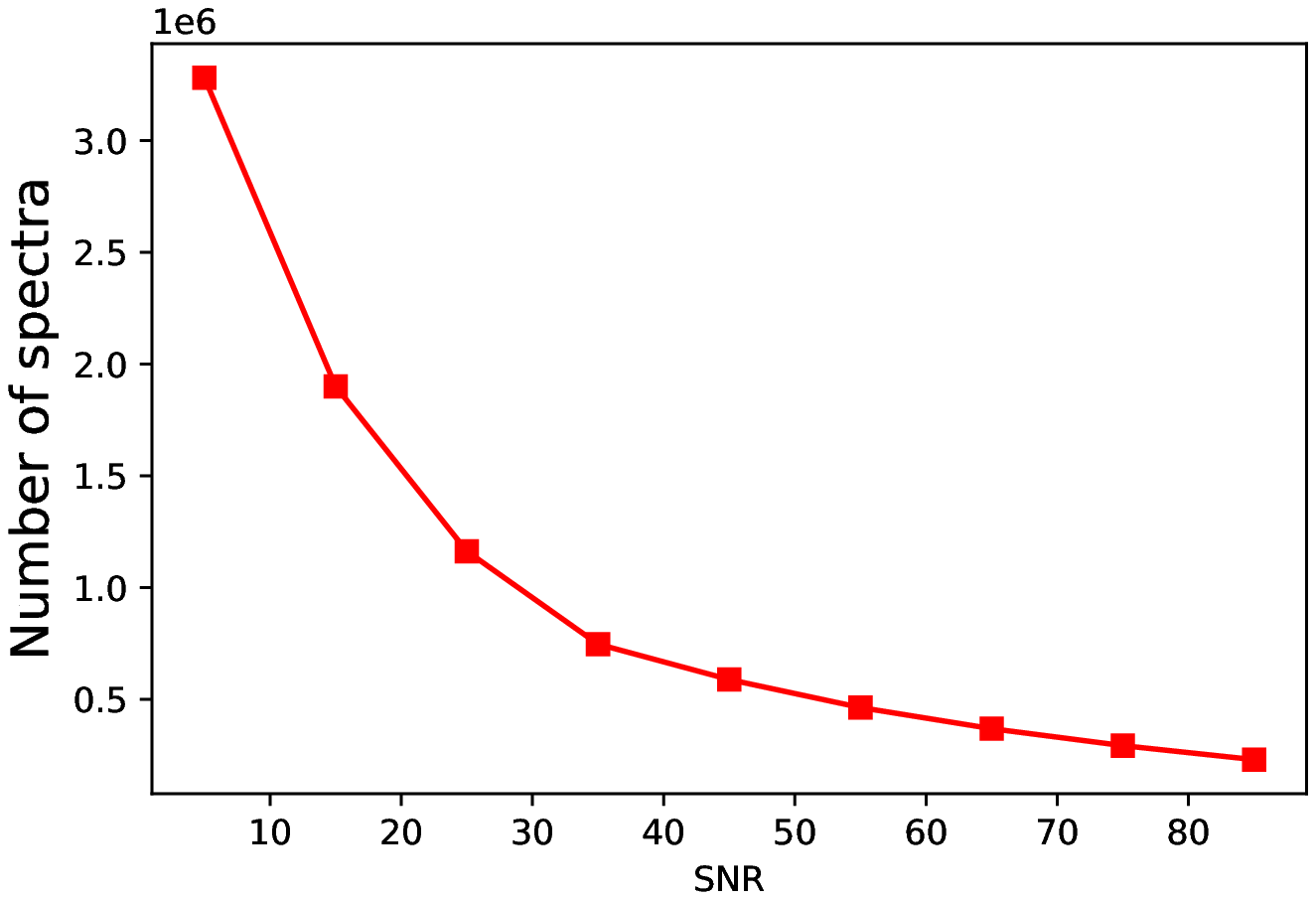}
 \includegraphics[width=7cm]{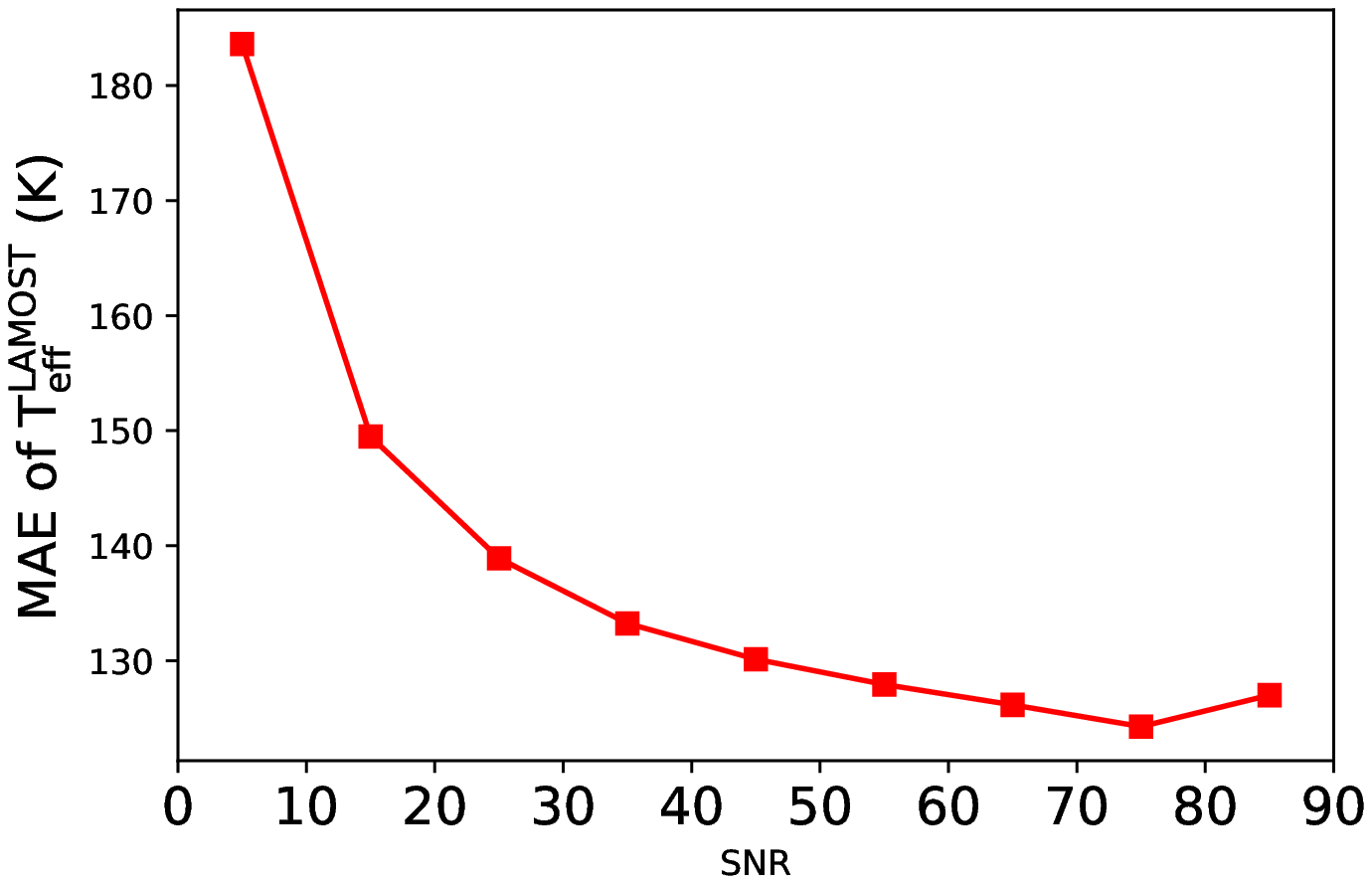}\\
 \includegraphics[width=7cm]{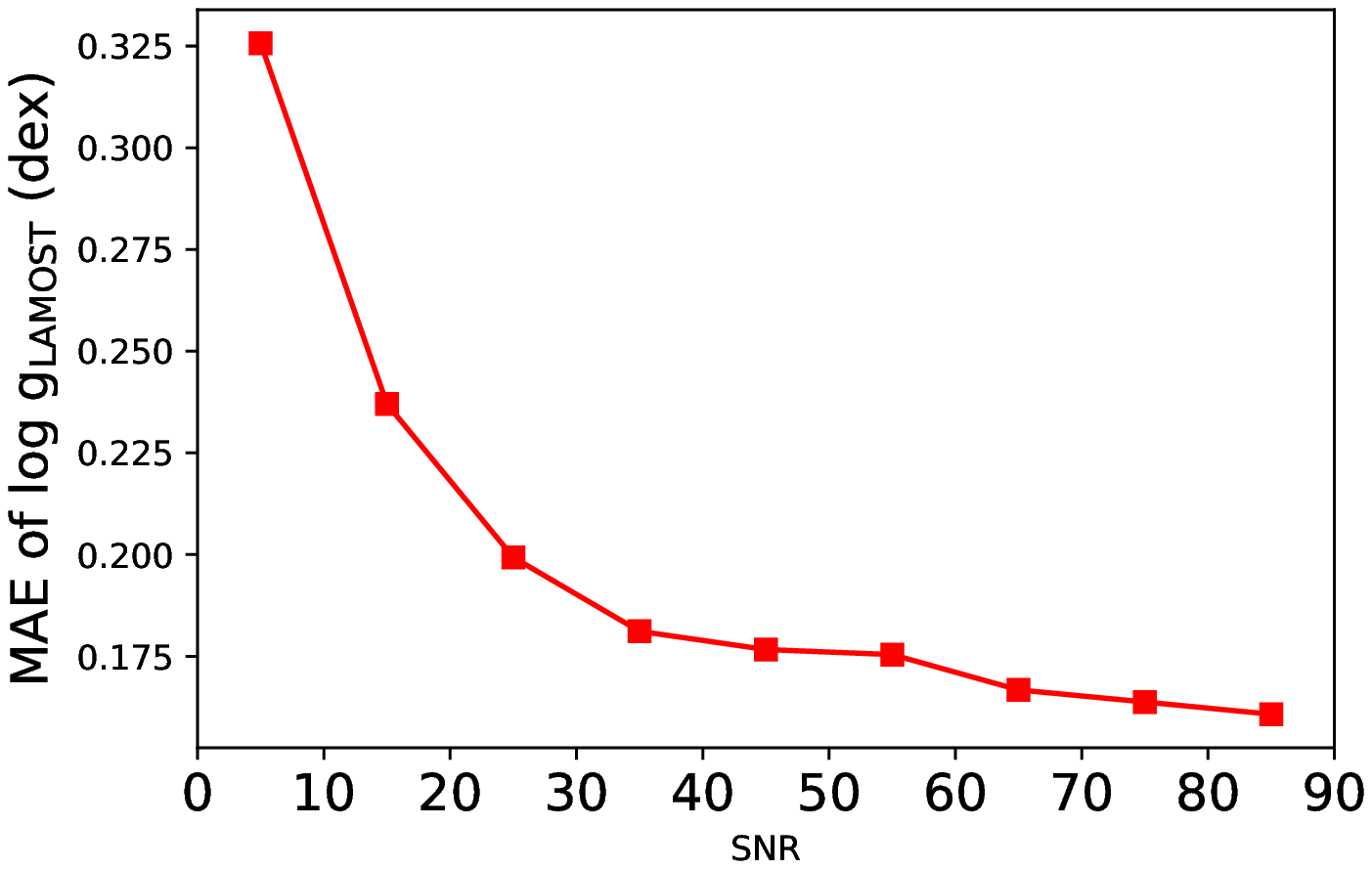}
 \includegraphics[width=7cm]{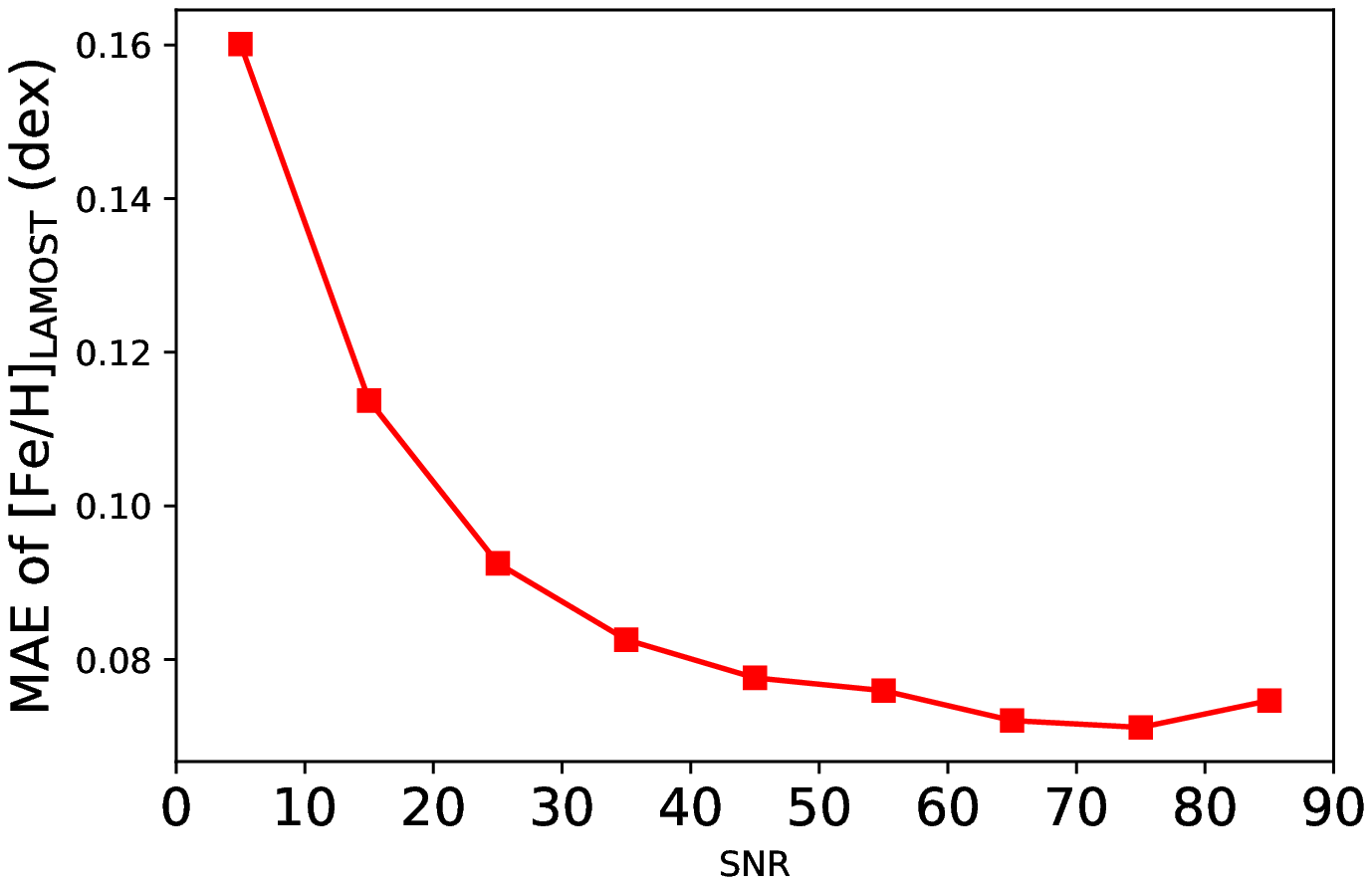}
\caption{The SNR characteristics of LAMOST stellar spectra and the dependencies of parameters estimation accuracy on signal-to-noise ratio (SNR). The parameter estimation performance is measured using the inconsistency between APOGEE and LAMOST pipeline on stellar spectra from their common stars.}
\label{fig:mae_change}
\end{figure}

The high signal-to-noise ratio (SNR) spectra contain less noise, and their spectral characteristics are obvious. Good results have been obtained for estimating stellar atmospheric parameters from the LAMOST high-SNR spectra. Unfortunately, the low-SNR spectra contain a lot of noise, their spectral characteristics are indistinguishable. Therefore, it is difficult to extract effective spectral features from them, which result in evidently degraded estimation performance. Comparing stellar atmospheric parameters provided by LAMOST DR8 with those provided by APOGEE DR12 (Fig.~\ref{fig:mae_change}), the inconsistencies increase sharply as the SNR decreasing, especially in the case of SNR $<$ 30. This phenomenon indicates that it is difficult to estimate stellar atmospheric parameters from low-SNR spectra.  Furthermore, the spectra with low SNR account for a large proportion in the LAMOST data. In LAMOST DR8, more than 60\% of the spectra are SNR $<$ 30 (Fig.~\ref{fig:mae_change}).
Therefore, it is potentially helpful to investigate more accurate methods for estimating stellar atmospheric parameters from low-SNR LAMOST spectra.

The difficulty in estimating stellar atmospheric parameters from low-SNR spectra lies in the feature extraction procedure. \cite{2015MNRAS.447..256B} investigated the spectral feature extraction problem based on principal component analysis (PCA).
\cite{2017MNRAS.464.3657X} extracted spectral features for stellar parameter estimation based on kernel principal component analysis (KPCA). Both PCA and KPCA are of global dimension reduction method which is sensitive to local noises and distortions. \cite{2014ApJ...790..105L} studied the spectral feature extraction problem based on LASSO (Least Absolute Shrinkage and Selection Operator) and local smoothing techniques. It is shown that the local dimension reduction method is effective in selecting the features for low signal-to-noise ratio spectra. Therefore, this paper uses the local feature extraction method LASSO to select features from LAMOST DR8 low-resolution spectra with 20$\leq$SNR$<$30.

After feature selection, we train an approximate model to learn a mapping from spectral features to a stellar atmospheric parameter, for example, $T_\texttt{eff}, \log~g$ and [Fe/H]. \cite{2015MNRAS.447..256B} used Gaussian Process Regression (GPR) to estimate stellar atmospheric parameters from SDSS DR10 spectra. \cite{2017MNRAS.464.3657X} used multiple-linear regression method to estimate stellar atmospheric parameters from LAMOST spectra. Unfortunately, for low-resolution and low-SNR spectra, this mapping is complex, and it is difficult to fit only by a basic non-linear regression model.
Fortunately, a series of works in literature show the effectiveness of neural networks (including deep learning) in estimating stellar atmospheric parameters from spectra \cite{2010PASP..122..608M,2014ApJ...790..105L,2017RAA....17...36L}. For example, \cite{2010PASP..122..608M} used artificial neural network (ANN) to estimate stellar atmospheric parameters from Gaia spectra with 5$<$SNR$<$25.
Therefore, this work use a special neural network, multilayer perceptron (MLP), to estimate stellar atmospheric parameters from LAMOST DR8 low-resolution spectra with 20$\leq$SNR$<$30.

The structure of this paper is as follows: Section \ref{sec:Data} introduced the spectra used to train and test the LASSO-MLP model, gave the pre-processing procedures for the spectra. Section \ref{sec:DimensionReductionLASSO} described the dimension reduction method LASSO. Section \ref{sec:ModelNN} described the MLP model. Section \ref{sec:APOGEE} verified the results of the LASSO-MLP model. Section \ref{sec:metal-poor} trained an ensemble LASSO-MLP model to estimate the [Fe/H] from the spectra of metal-poor stars. Finally, we summarized in section \ref{sec:future}. The estimation catalog, learned model, experimental
code, trained model, training data and test data are released on the following websit for scientific exploration and algorithm study: https://github.com/xrli/LASSO-MLP.
\section{Data and its pre-processing}\label{sec:Data}
\begin{figure}
 \includegraphics[width=4.8cm]{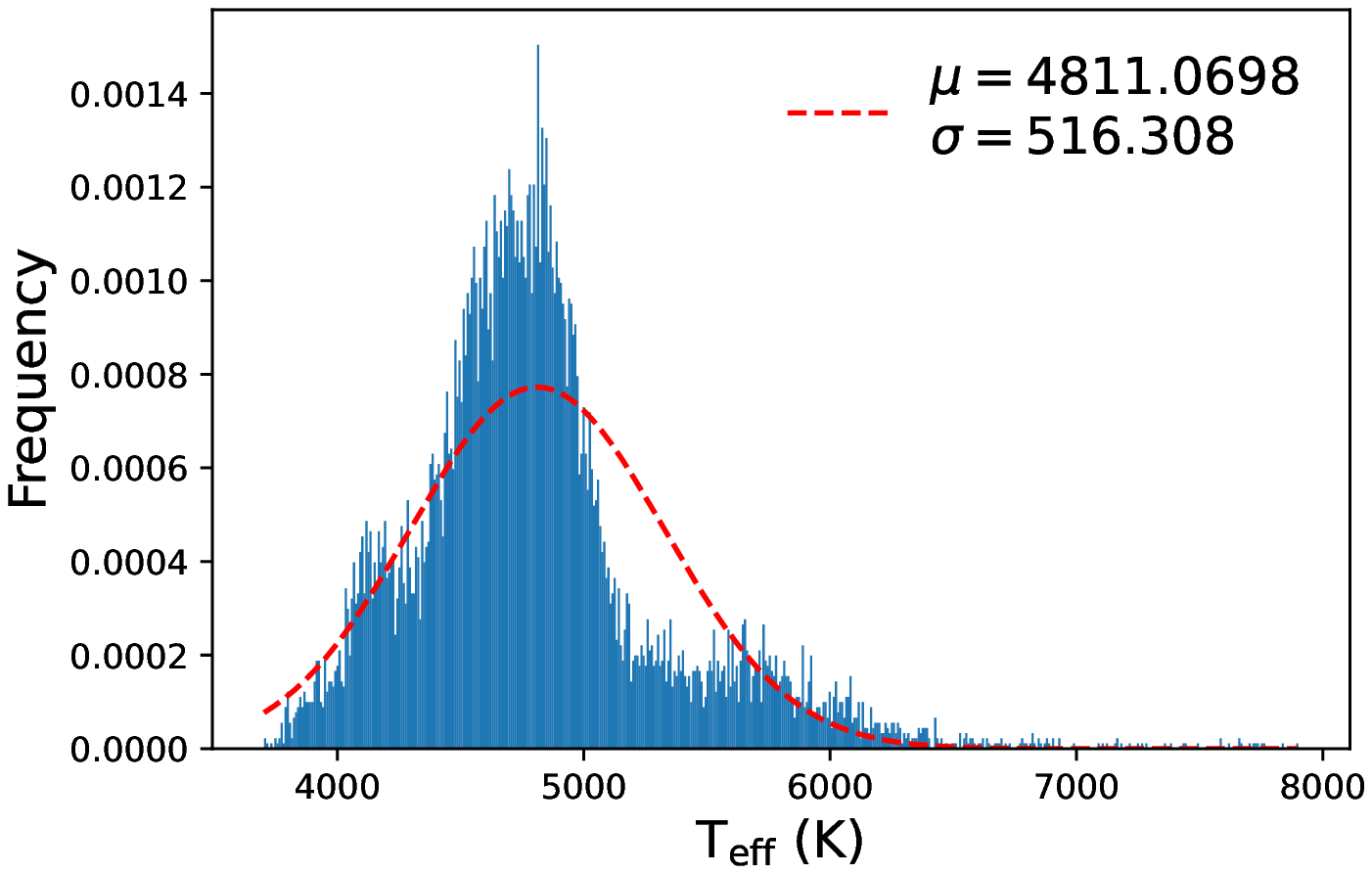}
 \includegraphics[width=4.8cm]{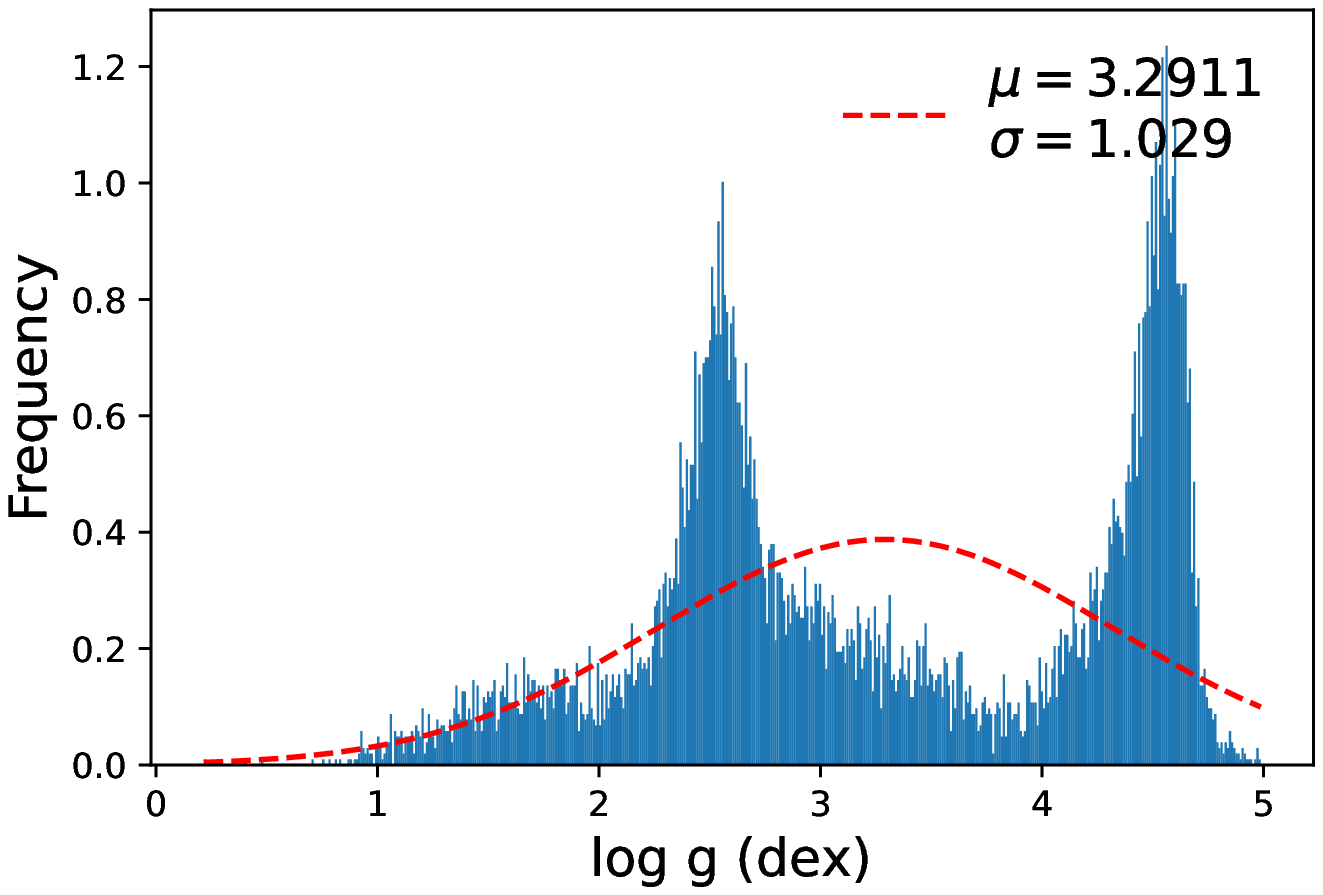}
 \includegraphics[width=4.8cm]{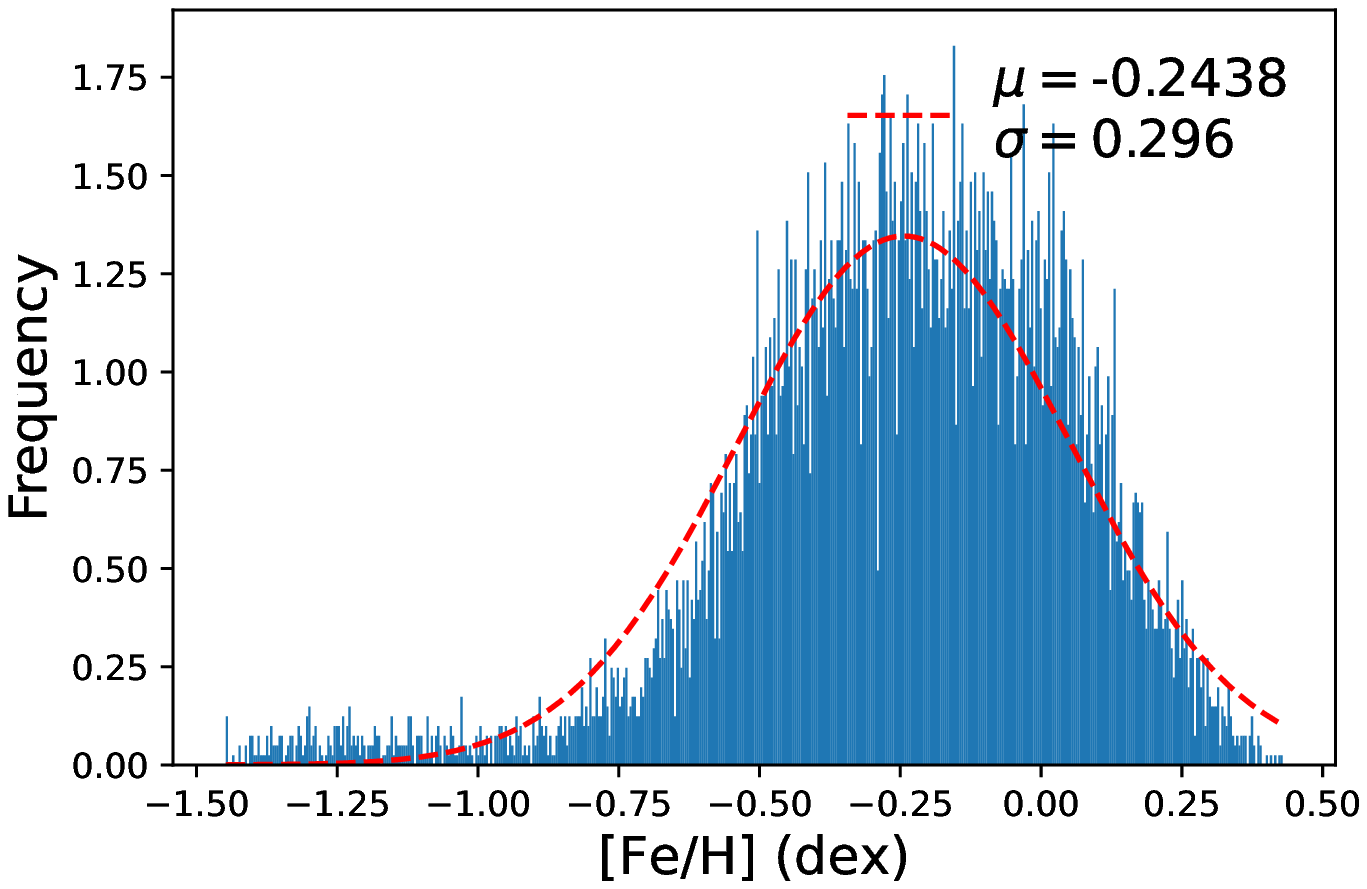}
\caption{Distribution of the reference spectral set.}
\label{fig:distribution}
\end{figure}

This work is to design a scheme for estimating the stellar atmospheric parameters from LAMOST low-resolution spectra with $20\leq$ SNR $<$30. The reference data consist of LAMOST spectra and their reference labels. The reference label is the parameters to be estimated, e.g., $T_\texttt{eff}, \log~g$ or [Fe/H]. Each of the LAMOST spectra in reference set has a unique common source observation in APOGEE high-resolution observations. The label comes from the APOGEE catalog estimated using ASPCAP from the common source high-resolution APOGEE observation. The ASPCAP (\cite{2016AJ....151..144G}) is a pipeline for estimating the stellar parameters from the APOGEE high-resolution spectra.
The common source matching is conducted based on the longitude-latitude constraint with a threshold 3.0 arc seconds.  If a LAMOST spectrum has multiple matching observation sources in APOGEE, then we remove the spectrum from the referene set.
Finally, the matched reference data set consists of 10,773 stellar spectra with $20\leq$ SNR $<$30 from common stars between APOGEE and LAMOST.
The ranges of the three stellar atmospheric parameters of these stars are $[3702, 7900]$ K for $T_{\texttt{eff}}$, $[0.216, 4.987]$ dex for $\log~g$, $[-1.448, 0.429]$ dex for [Fe/H]. Fig.~\ref{fig:distribution} shows the distribution of stellar atmospheric parameters. The reference data set are randomly divided into a training set and a test set at a ratio of 8:2.

\subsection{Pre-processings of the spectra}\label{sec:preprocessing}
The observed spectra are affected by the radial velocity and flux calibration. The radial velocity result in wavelength shift comparing with the theoretical spectra. Each of the above-mentioned factors can increase the difficulty of parameter estimation and reduce their accuracy. Therefore, it is necessary to do some pre-processing procedures to eliminate or reduce the potential negative  impacts from them.

The pre-processing procedures are as follows:
\begin{itemize}
\item Transform the observed spectra to their rest frame based on the radial velocity estimated by the LAMOST pipeline.
\item Cut the observed spectra based on their common wavelength range in the rest frame and resample them with a step 0.0001 in logarithmic wavelength coordinate system. The flux is linearly interpolated from the observed spectrum on the resampled wavelength. The computed spectrum of this step is denoted as $\boldsymbol f$. In this work, the common wavelength range is [3839.5, 8936.7]\AA.
\item Estimate the continuum. Firstly, a spectrum is processed using a median filtering algorithm to remove the spurious noises and spectral lines. The size of the filtering window is three pixels in the median filtering algorithm. Secondly, the continuum is estimated using a sixth-order polynomial fitting method. The estimated continuum is denoted as $\boldsymbol f_0$.
\item Divide the linear interpolated flux $\boldsymbol f$ by the fitted continuum $\boldsymbol f_0$ to normalize the spectra.
\end{itemize}

\begin{figure}
\center
\setlength{\belowcaptionskip}{-0.3cm}
\includegraphics[width=10cm]{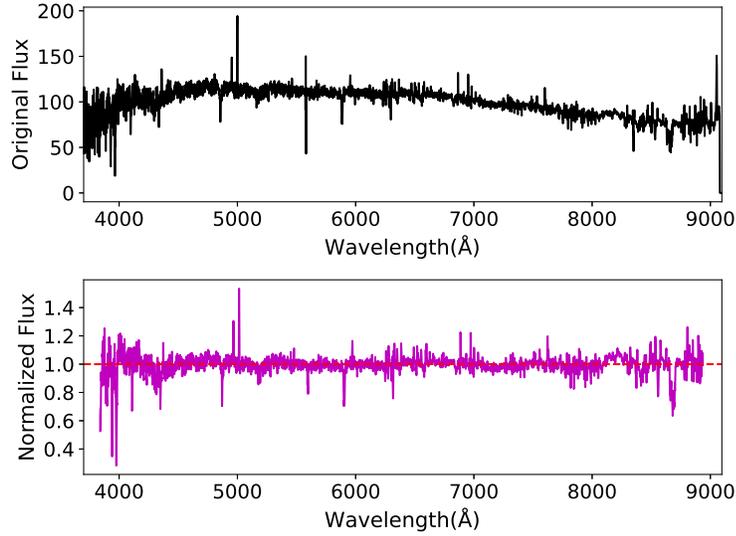}
\caption{Performance of spectrum pre-processing (Section. \ref{sec:preprocessing}). The above figure presents an observed spectrum and the below figure shows the result after pre-processing. }
\label{fig:preprocessing}
\end{figure}

An example of the above pre-processing is presented in Fig.~\ref{fig:preprocessing}.
\section{Dimension reduction for the spectra}\label{sec:DimensionReductionLASSO}
The preprocessed spectrum is a vector in a 3670 dimensional space, and there are a lot of noises and redundant components in this kind spectra. The noises and redundancies often lead to masking effects and accuracy degradation of parameter estimation. Although some spectral features are evident and sensitive to stellar atmospheric parameters on high quality spectra, their shape and contributions cannot be found in existence of serious noises by the stellar atmospheric parameter estimation model. This phenomenon is referred to as masking effects. Therefore, we need to reduce the dimension of these spectra to reduce ineffective or irrelevant components. By doing these, we can reduce the computational complexity of the model and the influence due to noises on the parameter estimation.

Suppose a vector $\boldsymbol x = \left(x_{1}, \cdots, x_{m}\right)^T$ represents a spectrum, $y$ represents a stellar atmospheric parameter $T_{\texttt {eff}}$, $\log~g$, or [Fe/H] of the corresponding spectrum, $T=\{(\boldsymbol{x}, y)\}$ is the training set, $\boldsymbol \beta = \left(\beta_{1}, \cdots, \beta_{m}\right)^T$ represents the weights that the model needs to learn. The objective function of LASSO is:
\begin{equation}
\hat{\boldsymbol{\beta}}=\arg \min _{\boldsymbol{\beta}}\left\{\sum_{(\boldsymbol x, y) \in T}\left(y-\sum_{i=1}^{m} \beta_{i} x_{i}\right)^{2}+\alpha \sum_{i=1}^{m}|\beta_{i}|\right\},
\end{equation}
where $\alpha$ is a preset parameter greater than 0, which controls the number of selected features. The $\beta_i$ with value zero indicates that the corresponding spectral flux $x_i$ is an irrelevant component or redundancy component. Otherwise, a non-zero $\beta_j$ indicates that the corresponding $x_j$ is a useful component for stellar atmospheric parameter estimation.

To further explore the optimality of LASSO in dimension reduction for the spectra from LAMOST DR8 low-resolution with $20\leq$ SNR $<30$, we compared linear dimension reduction methods with non-linear dimension reduction methods, local dimension reduction methods with global dimension reduction methods. This paper explored PCA (Principal Component Analysis), KPCA (Kernel Principal Component Analysis), ISOMAP (Isometric Mapping), MDS (Multidimensional Scaling), LLE (Locally Linear Embedding), and LASSO. The PCA is a linear global dimension reduction method, LASSO is a local linear dimension reduction method, KPCA, ISOMAP, and MDS are nonlinear global dimension reduction methods, LLE is a local manifold dimension reduction method.

We compared the scheme without dimension reduction (the first row of the Table~\ref{tab:reduction}) with schemes based on dimension reduction (the second to seventh rows of the Table~\ref{tab:reduction}). It is shown that the schemes based on PCA, KPCA, and LASSO are better than the scheme without dimension reduction. This phenomenon indicates the existence of the sparseness in the high-dimensional spectral space. Therefore, it is difficult for the model to find the data characteristics of the samples, and parameter estimation without dimension reduction will reduce the efficiency and accuracy of the model. These results indicate the necessity of dimension reduction in estimating stellar atmospheric parameters from spectra.

We compared the linear dimension reduction methods PCA, LASSO (the second and seventh rows of the Table~\ref{tab:reduction}) with the non-linear dimension reduction methods KPCA, ISOMAP, LLE, MDS (the third to sixth rows of the Table~\ref{tab:reduction} ). It is shown that the estimation from the linear dimension reduction methods are better than the nonlinear dimension reduction methods. The used model in this work can be represented by X-MLP, where X represents a  dimension reduction method. These results indicate that the linear dimension reduction methods are more suitable for the stellar atmospheric parameter estimation scheme of X-MLP model.

We compared the global dimension reduction methods PCA and KPCA (the second and third rows of the Table~\ref{tab:reduction}) with the local dimension reduction methods LLE, LASSO (the fifth and seventh rows of the Table~\ref{tab:reduction}). It is shown that the estimation of the local linear method LASSO are better than the local non-linear method LLE and the global dimension reduction methods PCA, KPCA. A feature calculated by the global dimension reduction method uses almost all the observed pixels. This characteristics make each extracted features can be afftected by any noise and distortion; Conversely, a feature of the local dimension reduction method is computed only using a small subset of observed fluxes, for example, several fluxes near a spectral line. Furthermore, the LASSO can adaptively discard some ineffective pixels according to the overall balance between the effects from noises, distortions and spectral characteristics. Therefore, the features extracted by the local method may be less affected by noise and distortion. These experimental results indicate that the parameter estimation scheme X-MLP based on a global dimension reduction is less robust, and the parameter estimation method based on a local dimension reduction performs good, especially the LASSO-MLP method with the property of discrimination and rejection.

Based on the above-mentioned studies, it is shown that the local dimension reduction methods and the linear dimension reduction methods are more suitable for the stellar atmospheric parameter estimation. Therefore, we used the local linear dimension reduction method LASSO to reduce the dimension for the spectra from LAMOST DR8 low-resolution with $20\leq$ SNR $<30$.

\begin{table}
\renewcommand\arraystretch{1.5}
\caption{Comparisons between linear dimension reduction methods and non-linear dimension reduction methods, local dimension reduction methods and global dimension reduction methods.}
\label{tab:reduction}
\centering
\begin{tabular}{|c|c|c|c|c|c|c|c|c|c|}
\hline
 & \multicolumn {3}{c|}  {$T_{\texttt{eff}}$} & \multicolumn {3}{c|}{$\log~g$} & \multicolumn {3}{c|}{[Fe/H]}\\
\cline{2-10}
& MAE (K) & $\mu$ & $\sigma$ & MAE (dex) & $\mu$ & $\sigma$ & MAE (dex) & $\mu$ & $\sigma$ \\
\hline
MLP & 107.5 & -1.750& 203.5& 0.611&0.343& 6.322&0.128 & 0.0070 & 0.566 \\ \hline
PCA-MLP &100.1 &-0.645 &2006.1 &0.191 &-0.017 &0.726 &0.062 &0.0012 &0.164 \\ \hline
KPCA-MLP & 91.57 & 0.118 &175.6 &0.686 &-0.011 &0.424 &0.071 &-0.0072 &0.426 \\ \hline
ISOMAP-MLP &243.4 &-0.0046 & 397.9&0.619 &0.100 &3.080 & 3.680 &3.646 &3.185 \\ \hline
LLE-MLP &127.3 &-5.108 &196.60 &0.286 &-0.019 &0.452&3.526 &3.526 &0.968 \\ \hline
MDS-MLP & 454.0 & -230.2 & 554.8 & 1.284 & 0.069 & 1.801 & 0.300 & -0.109 & 1.006\\ \hline
LASSO-MLP  &\textbf{84.32} & \textbf{0.205} & \textbf{164.8} & \textbf{0.137} & \textbf{-0.00084} & \textbf{0.217} & \textbf{0.063} & \textbf{-0.00035} & \textbf{0.095} \\ \hline
\end{tabular}
\end{table}
\section{The stellar atmospheric parameter estimation model}\label{sec:ModelNN}
After dimension reduction, the information of every spectrum can be represented by a vector by stacking the selected features. In estimating $T_\texttt{eff}, \log~g$, or [Fe/H], the information of a spectrum can respectively represented by a 141, 553, and 833-dimension vector. From this feature vector, we can estimate the stellar atmospheric parameter using a regression method. This work estimated the stellar atmospheric parameters by a MLP.

A MLP provides a global non-linear mapping from input (a spectrum $\boldsymbol x$) to an output (the stellar atmospheric parameter of the corresponding star, $y =$ $\log~T_\texttt{eff}, \log~g$, or [Fe/H]). Each node in a layer of the MLP is fully connected with the nodes in its previous layer. The first layer is referred to as input layer, the middle one is the hidden layer, and the last one is the output layer. Except for the input node, each node is a neuron with a non-linear activation function.

Suppose $S=\{(\mathbf{x}_i, y_i), i=1,\cdots,N\}$ is a set of training data, where $N$ is the number of stellar spectra used for learning a MLP model, $\mathbf{x}_i$ represents a stellar spectrum, and $y_i$ is the reference value of the atmospheric parameter $T_\texttt{eff}$, log~$g$ or [Fe/H] of the spectrum $\mathbf{x}_i$. In this work, the reference values of the stellar atmospheric parameters are estimated by the ASPCAP.  Let $h_{W, \mathbf{b}}(\mathbf{x}_i)$ denote the estimation of $y_i$ from the spectrum $\mathbf{x}_i$ using the MLP, $W$ and $\mathbf{b}$ respectively represent the sets of connection weights and biases in the MLP model. To learn the model parameters $W$ and $\mathbf{b}$, a mean squared error loss function can be used:
\begin{equation}
\operatorname{Loss}(S|W, \mathbf{b}) = \frac{1}{N}\sum\limits_{i=1}^N (h_{W, \mathbf{b}}(\mathbf{x}_i)- y_i)^{2}.
\end{equation}
The model parameters $W$ and $\mathbf{b}$ are optimized by iteratively minimizing the loss function. When iterations reach a preset maximum number of times or the loss function is smaller than a given threshold, we stop the iterations.

To evaluate the optimality of the MLP in estimating stellar atmospheric parameter from the LAMOST DR8 low-resolution spectra with $20\leq$ SNR $<30$, we investigated the performances of multiple typical regression methods. For example, LR (Linear Regression), ridge, LASSO, ElasticNet,SVR (Support Vector Regression), KNR (K-Neighbors Regression), DecisionTree, GradientBoosting, XGBoost, lightGBM, and Random forest. The ranges of the three stellar atmospheric parameters in this work are $[3594, 7900]$ K for $T_{\texttt{eff}}$, $[0.216, 4.987]$ dex for $\log~g$, $[-1.448, 0.429 ]$ dex for [Fe/H]. To increase the numerical computation performance, we used $\log~T_{\text {eff}}$ instead of $T_{\text {eff}}$. In addition, we standardized each selected feature to zero mean and one variance. This standardization helps to improve the stability of most machine learning algorithms.

\begin{table}
\renewcommand\arraystretch{1.5}
\caption{Comparisons between MLP and some typical regression methods.}
\label{tab:model}
\centering
\begin{tabular}{|c|c|c|c|c|c|c|c|c|c|}
\hline
 & \multicolumn {3}{c|}  {$T_{\texttt{eff}}$} & \multicolumn {3}{c|}{$\log~g$} & \multicolumn {3}{c|}{[Fe/H]}\\
\cline{2-10}
& MAE (K) & $\mu$ & $\sigma$ & MAE (dex) & $\mu$ & $\sigma$ & MAE (dex) & $\mu$ & $\sigma$ \\ \hline
LR & 156.3 & -14.31 & 223.9 & 0.188 & -0.013 &0.286 &0.092 &-0.0015 &0.146 \\ \hline
Ridge & 156.3 & -14.31 & 223.9 & 0.188 & -0.013 & 0.286 & 0.101 & 0.014 & 0.841 \\ \hline
Lasso & 157.9 & -15.03 & 226.7 & 0.188 & -0.013 & 0.285 & 0.084 & 0.00053 & 0.249 \\ \hline
ElasticNet  &156.3 &-14.31 &223.9 &0.192 &-0.013 &0.291 &0.086 &0.003 &0.332 \\ \hline
SVR & 132.3 & -1.066 & 190.11 & 0.174 & -0.003 & 0.273 & 0.070 & -0.002 & 0.098   \\ \hline
KNR  &139.9 &0.633 &197.9 &0.194 &0.0016 &0.289 &0.097 &0.0043 &0.133 \\ \hline
DecisionTree &173.2 &-0.223 &247.1 &0.299 & -0.010 &0.450 &0.156 &-0.0087 &0.211 \\ \hline
GradientBoosting &135.7 &1.643 &199.9 &0.189 &-0.010 &0.267 &0.084 &-0.0022 & 0.116\\ \hline
XGBoost & 130.1 & -4.954 & 185.4 & 0.191 & -0.0090 & 0.268 & 0.066 & 0.65 & 0.28 \\ \hline
lightGBM & 123.8 & -1.079 & 177.1 & 0.166 & -0.0070 & 0.239 & 0.072 & -0.0013 & 0.101 \\ \hline
Random forest & 122.8 & -0.293 & 177.0 & 0.188 & -0.011 & 0.273 & 0.093 & -0.0037 & 0.126 \\ \hline
MLP  &\textbf{84.32} & \textbf{0.205} & \textbf{164.8} & \textbf{0.137} & \textbf{-0.00084} & \textbf{0.217} & \textbf{0.063} & \textbf{-0.00035} & \textbf{0.095} \\ \hline
\end{tabular}
\end{table}

The LR is one of the most commonly used algorithms for processing regression tasks. However, the naive linear regression is usually replaced by the regularized regression methods (LASSO regression, Ridge regression and Elastic-Net). The Ridge regression is a linear regression with a L2 regularization, the LASSO regression is a linear regression method with a L1 regularization, and the Elastic-Net regression is a linear regression mothod combined with a L1 regularization and a L2 regularization. The high dimension of the spectra tends to result in over-fitting, and the regularization is actually a technique that penalizes too many regression coefficients to reduce the risk of over-fitting. We compared the ordinary linear regression method (the first row of the Table~\ref{tab:model}) , Ridge regression (the second row of the Table~\ref{tab:model}), LASSO regression (the third row of the Table~\ref{tab:model}), and Elastic-Net regression (the 4th row of the Table~\ref{tab:model}). It is shown that the regularization can indeed improve the performance of linear estimation method on [Fe/H]. However, these linear methods are inferior to the non-linear regression methods, which will be discussed furtherly in following paragraphs. These experimental results indicate that there exsit some non-linear relationships between the spectral features and the the stellar atmospheric parameters. Therefore, it is necessary to investigate the estimation performance of some typical non-linear regression methods.

The SVR, instance-based KNR, and DecisionTree are three typical non-linear regression methods, and their experimental results are presented in the 5th-7th rows of the Table~\ref{tab:model}. Although the dimension of the spectra is high, the SVR is robust to overfitting in a high-dimensional space. Therefore, the SVR achieves better performance than the linear estimation methods. Due to the number of reference spectra in the training set, the KNR can find more similar training spectra for a spectrum to be parameterized, and the experimental results indicates that KNR also outperforms the linear regression methods. However, the estimations from low-SNR spectra by DecisionTree is prone to overfitting, which leads to worse performance than the linear regression models. Therefore, we need to further investigate the ways to prevent overfitting in tree-based schemes.

The ensemble learning scheme helps to eliminate or reduce overfitting phenomenon. The ensemble learning increase the generalization ability and robustness of a model by combining the prediction results of multiple basic learners. According to the generation method of basic learners, the ensemble learning methods are roughly divided into two categories: In the first category, there are strong dependencies between the basic learners, which must be generated successively; In the second category, the basic learners are independent from each other, and they can be generated parallelly and independently. The Gradient Boosting, XGBoost and lightGBM are the representatives of the first category, and Random Forest is the representative of the second category. The experimental results of the above-mentioned ensemble learning methods are presented in the 8th-11th rows of the Table~\ref{tab:model}.

The basic idea of Gradient Boosting is to train the newly added weak classifier according to the negative gradient information of the current model loss function, and integrate the trained weak classifiers into the existing model in the form of accumulation. This process is to continuously reduce the loss function and the model deviation. Due to the excessive pursuit of reducing errors, Gradient Boosting is prone to overfitting, and takes a long time to train. Experimental results show that Gradient Boosting is inferior to the other ensemble learning methods in estimating the stellar atmospheric parameters (the 8th row of the Table~\ref{tab:model}). Therefore, the XGBoost adds a regularization term to the cost function to improve generalization ability by controling the complexity of the model. From the perspective of balancing variance and bias, it reduces the variance of the model, makes the learned model simpler, and reduces the risk of overfitting. Experimental results show that the XGBoost outperforms the Gradient Boosting in estimating stellar atmospheric parameters (the 8th and 9th rows of the Table~\ref{tab:model}). The lightGBM mainly optimizes the training speed of the model, and its basic principle is similar to the XGBboost. Therefore, there is not essential difference on accuracy between these two methods (the 9th and 10th rows of the Table~\ref{tab:model}). On the basis of building an ensemble with decision tree as the basic learner, Random forest further introduces random feature selection in the training process. This randomness makes the model have more generalization ability. Experimental results show that the Random forest outperforms decision tree in estimating the stellar atmospheric parameters (the 7th and 11th rows of the Table~\ref{tab:model})

Random forest consists of multiple decision trees which is independent from each other. On the other hand, the MLP consists of multiple layers where each layer is fully with the layer before it. Experimental results show that Random forest is inferior to MLP in estimating the stellar atmospheric parameters (the 11th and 12th rows of the Table~\ref{tab:model}). Therefore, we used MLP to estimate the stellar atmospheric parameters from LAMOST low-resolution spectra with $20\leq$ SNR $<30$.

\section{Performance Evaluation of LASSO-MLP}
\label{sec:APOGEE}

We evaluated the reliability and accuracy of the proposed model LASSO-MLP from two aspects. Firstly, it is evaluated by computing the consistencies between the LASSO-MLP estimations and the APOGEE estimation from high-resolution spectra (the 1st row of the Table~\ref{tab:consistency}). Secondly, by treating the ASPACP estimations from APOGEE high-resolution spectra as benchmark, we compared statistical characteristics of the LASSO-MLP estimations and LASP estimations (the 1st and 2nd rows of the Table~\ref{tab:consistency}). These evaluations are conducted based on the following three statistical indicators: mean absolute error (MAE), mean error ($\mu$), and standard deviation of error ($\sigma$).  The MAE is the average of the absolute values of errors, which can avoid the problem of mutual cancellation from errors on various spectra, and can measure the overall accuracy of an estimation model. The $\mu$ is the arithmetic mean of the error, representing the most likely value of the error, reflecting the systematic bias of a parameter estimation model. The $\sigma$ describes the fluctuation around the average estimation, which reflects the uncertainty of a model.

\begin{figure}
 \includegraphics[width=5cm]{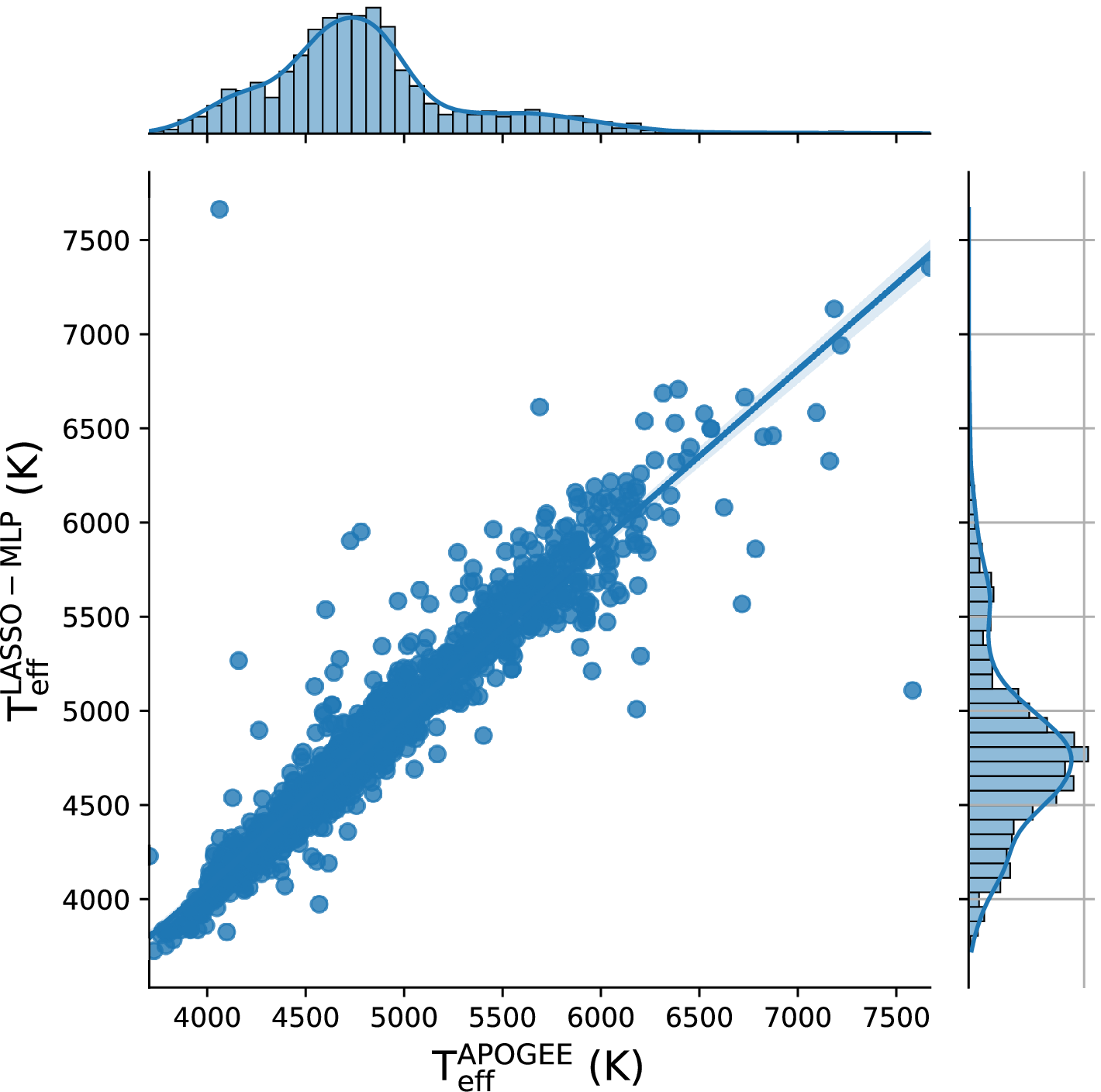}
 \includegraphics[width=5cm]{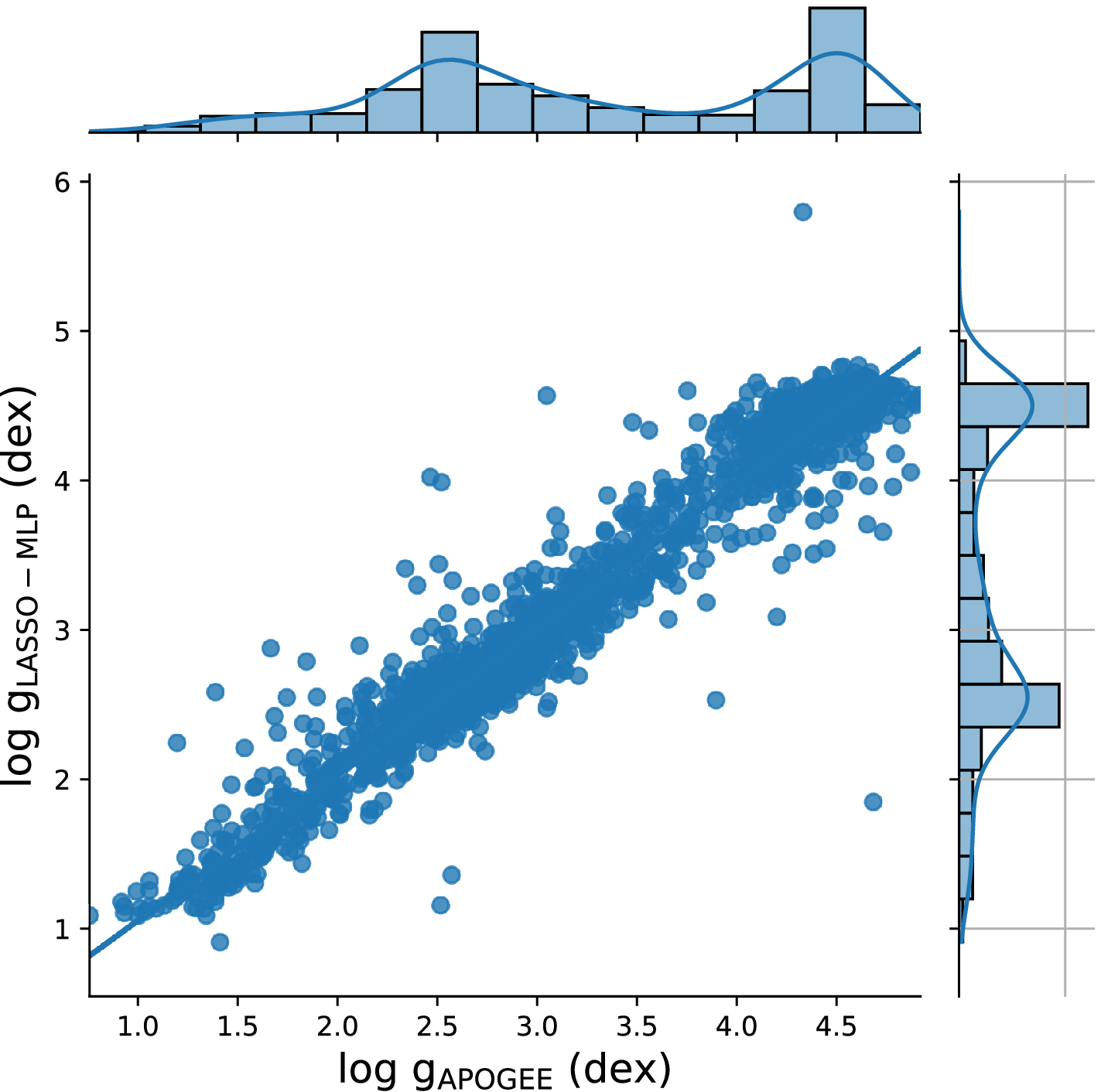}
 \includegraphics[width=5cm]{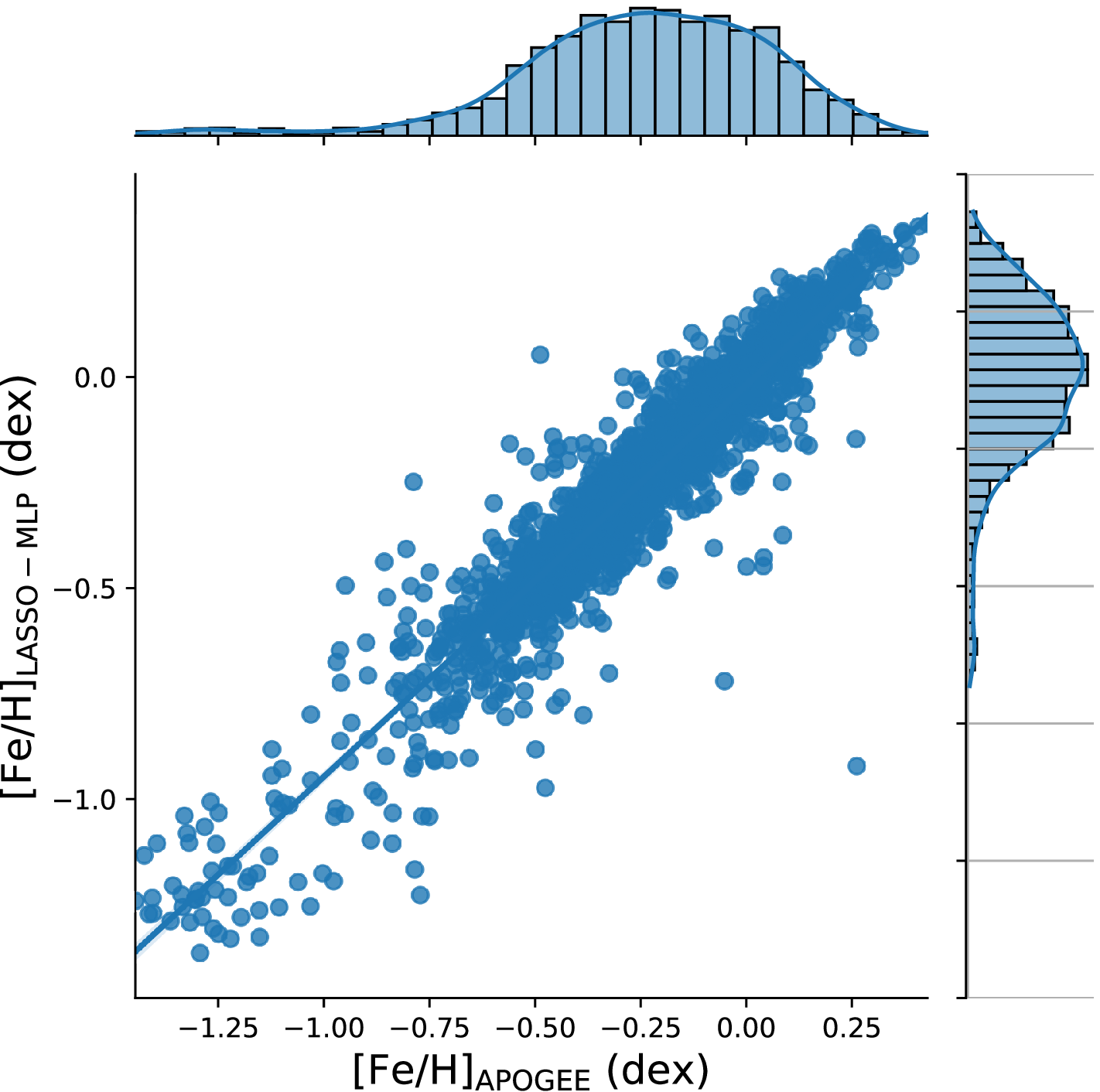}\\
 \includegraphics[width=5cm]{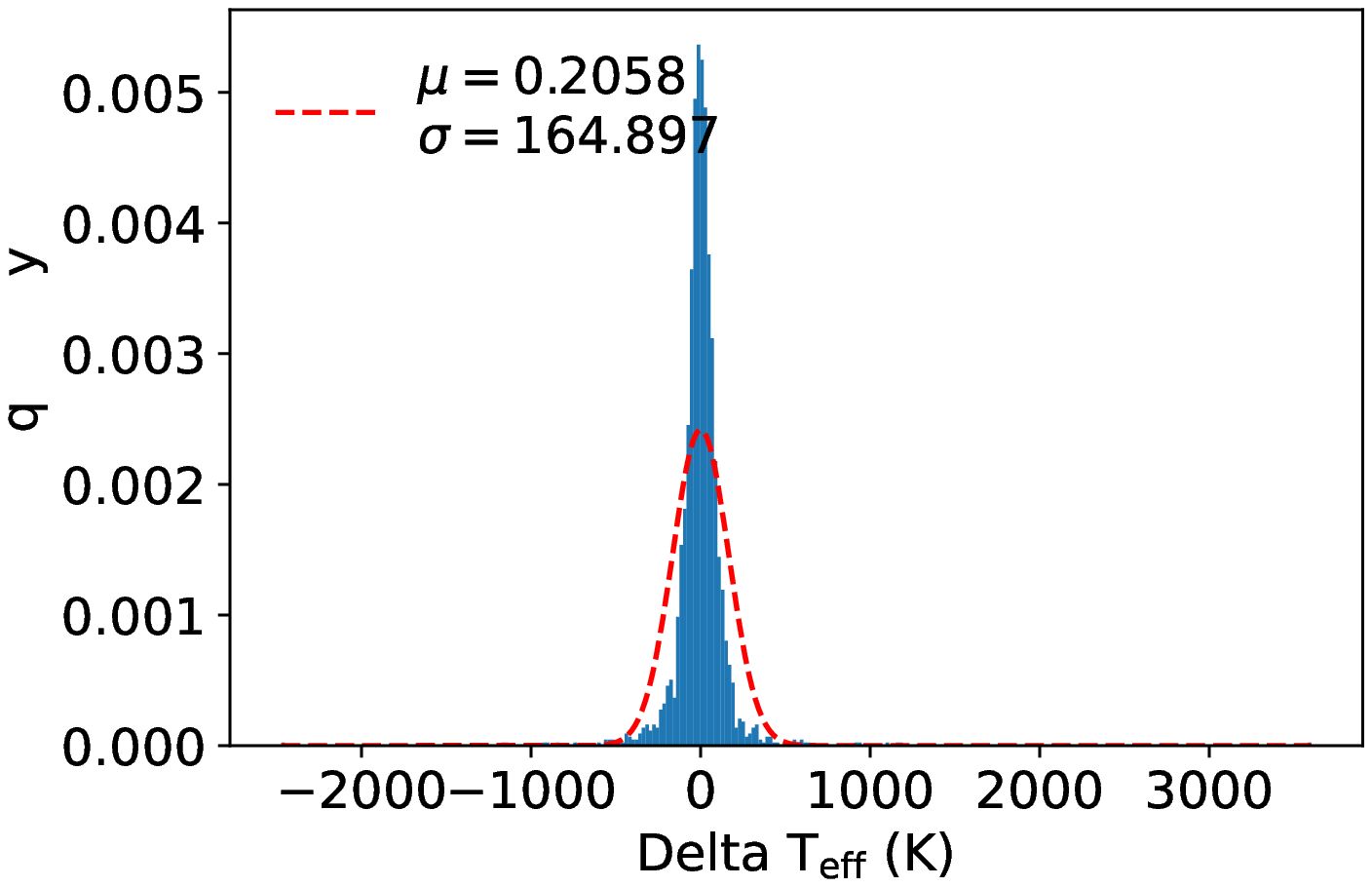}
 \includegraphics[width=5cm]{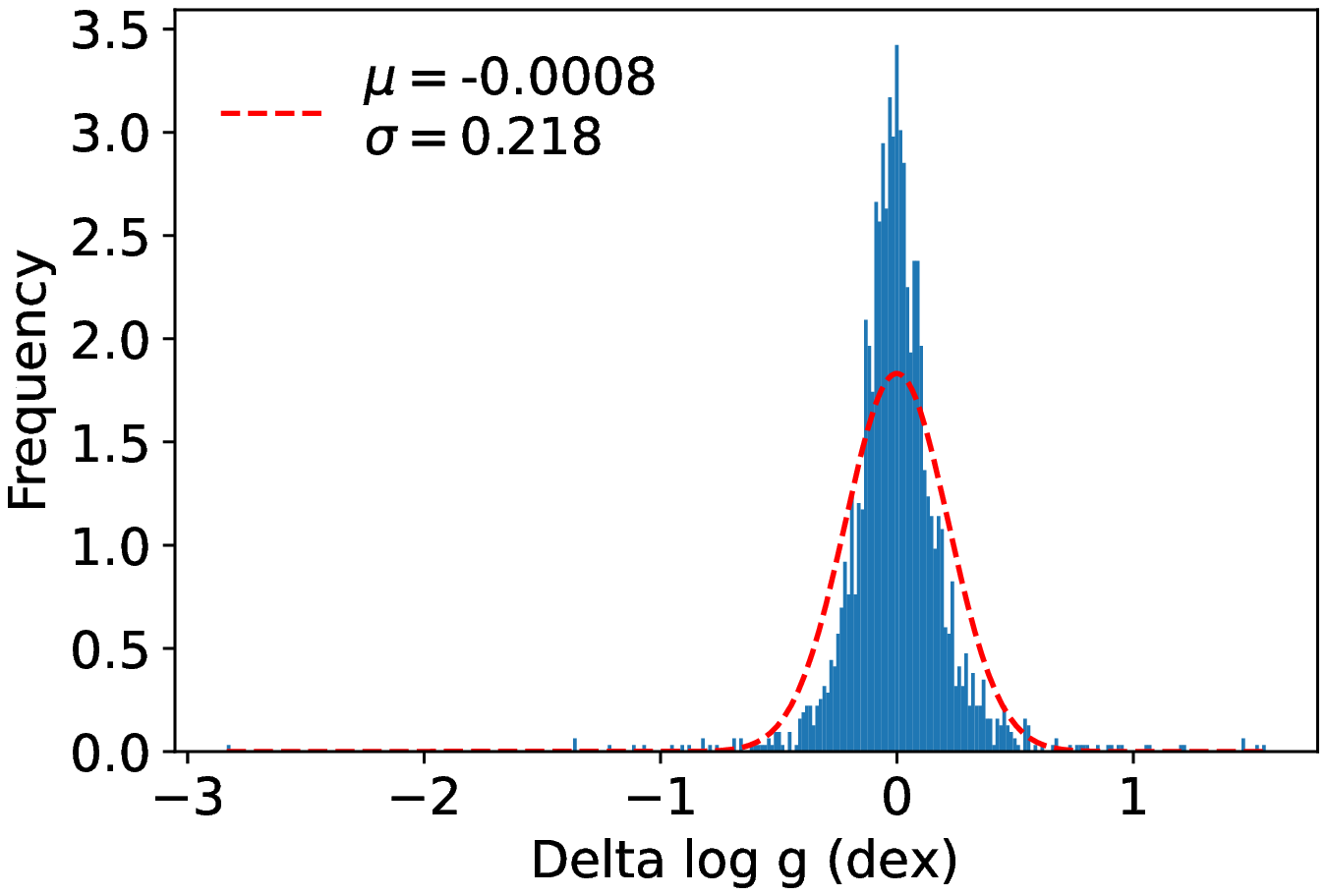}
 \includegraphics[width=5cm]{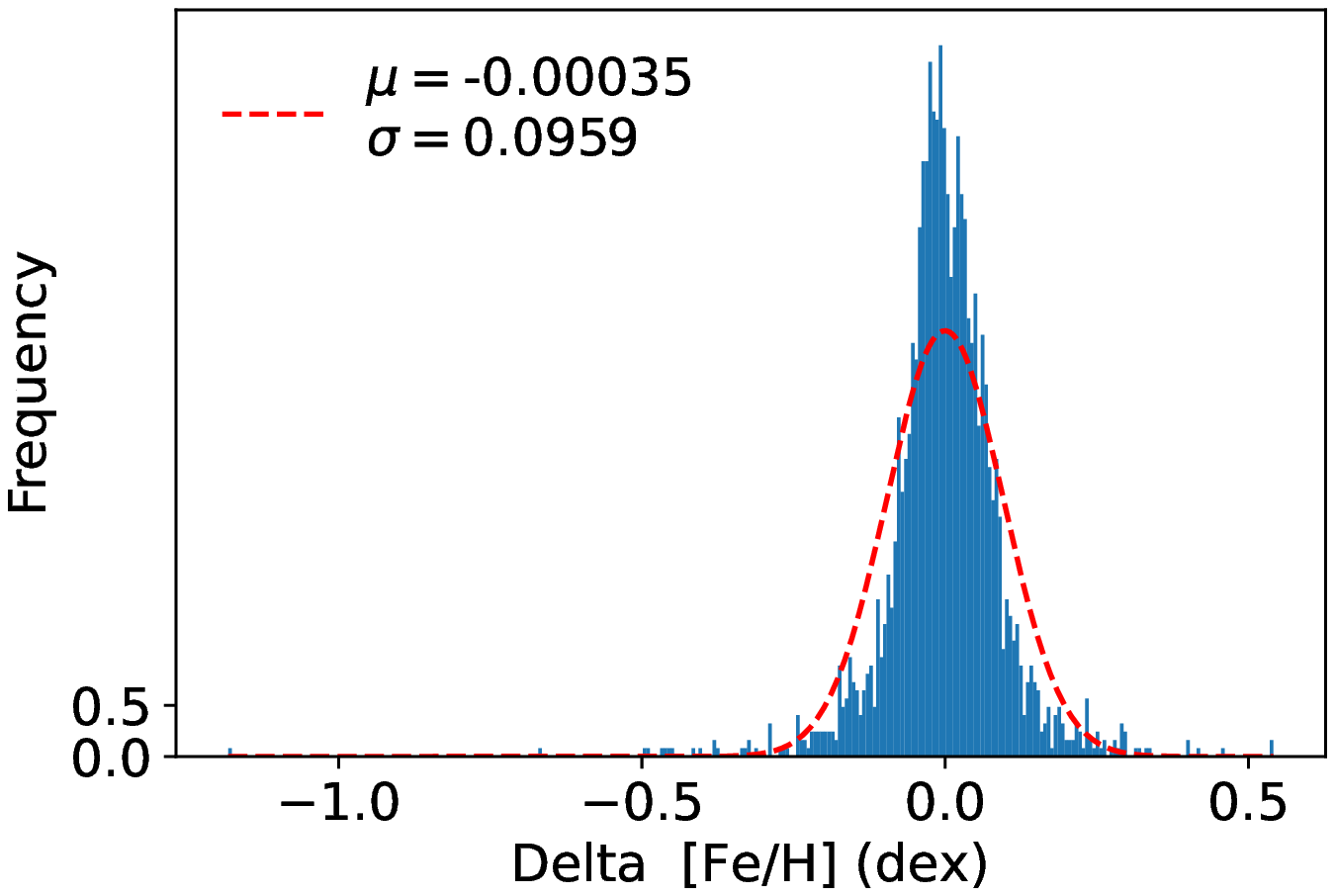}\\
\caption{Compare the estimated values of $T_{\texttt{eff}}$, $\log~g$ and [Fe/H] with the corresponding reference values provided by APOGEE. The above is the binary correlation diagram. The following figures are the histogram of the errors, The dotted line represents the Gaussian fitting to the residual distribution, and the top of the histogram shows the average ($\mu$) and standard deviation ($\sigma$) of the error.}
\label{fig:apogee_mlp}
\end{figure}

Three statistical indicators (MAE, $\mu$, $\sigma$) are all relatively small in the scenario of low-resolution and low-SNR spectra ( the first row of the Table~\ref{tab:consistency}). This result indicates an excellent consistency between LASSO-MLP estimations from LAMOST spectra and the ASPCAP estimations from APOGEE high-resolution spectra, and this consistency is stable on the whole.
At the same time, the LASSO-MLP estimation does not show any obvious systematic shift on various parameter intervals (Fig.~\ref{fig:apogee_mlp}). Therefore, the LASSO-MLP model has a strong generalization ability in estimating the stellar atmospheric parameters from the low-SNR spectra.

\begin{figure}
\includegraphics[width=5cm]{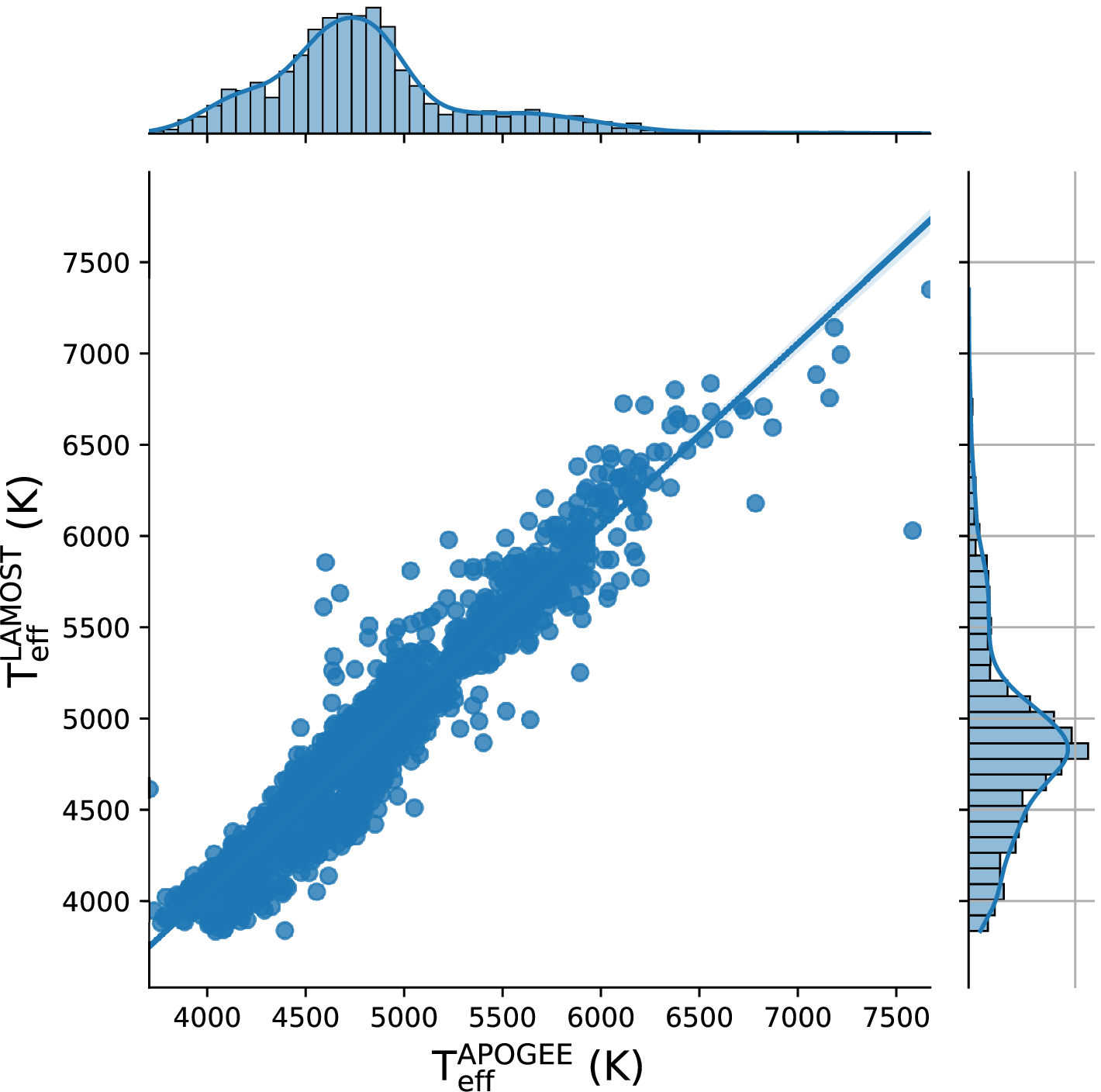}
\includegraphics[width=5cm]{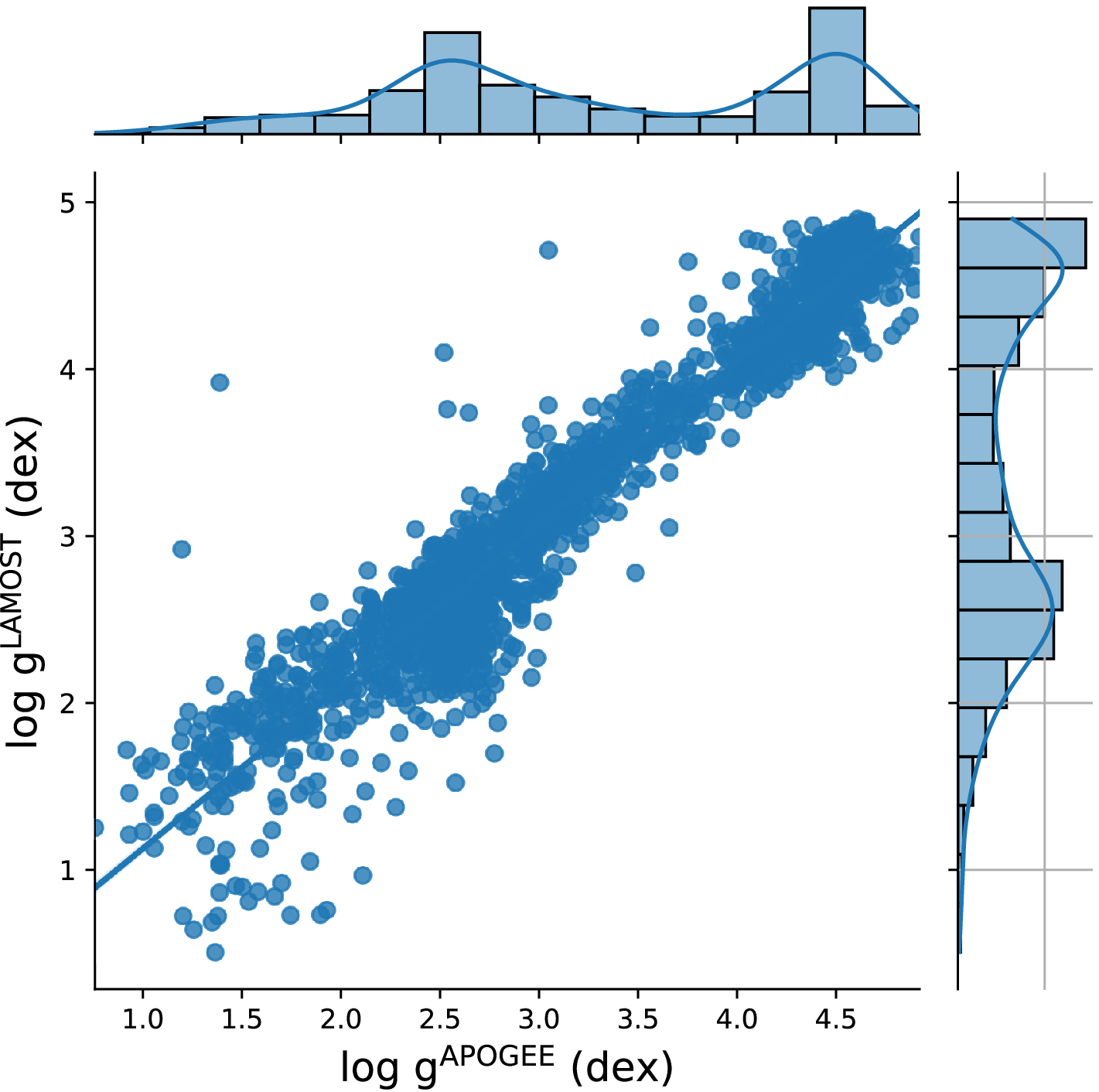}
\includegraphics[width=5cm]{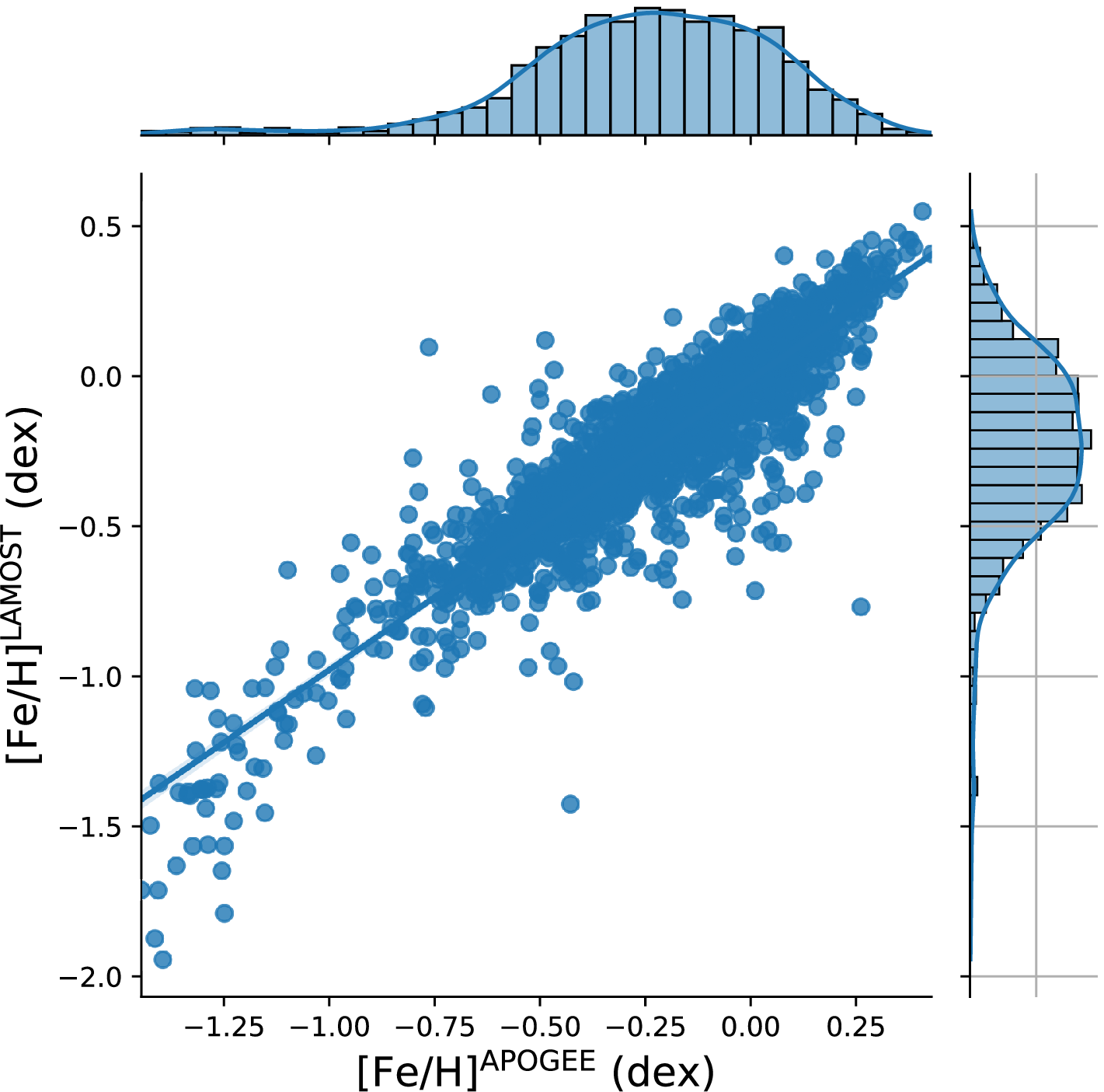}\\
\includegraphics[width=5cm]{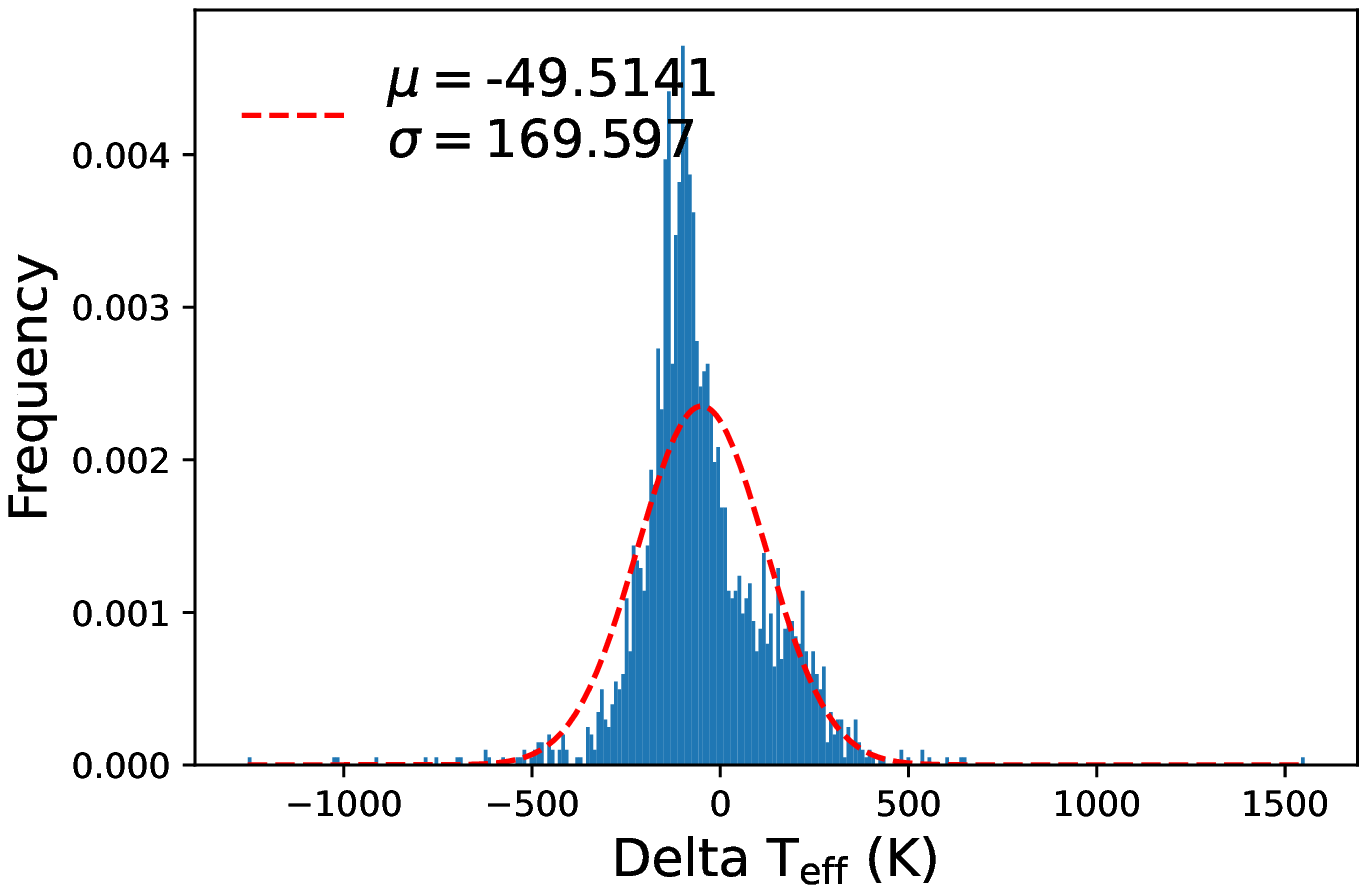}
\includegraphics[width=5cm]{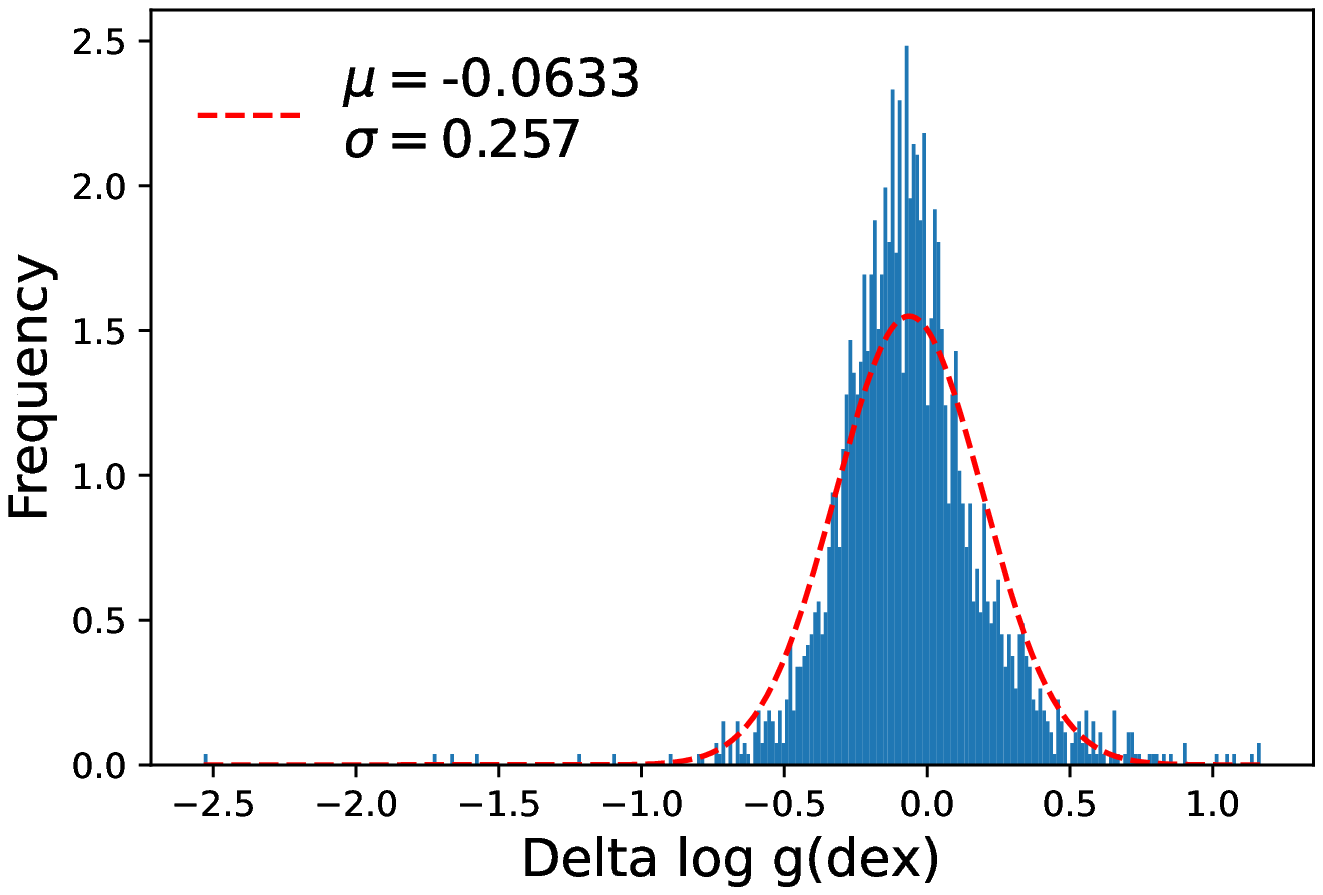}
\includegraphics[width=5cm]{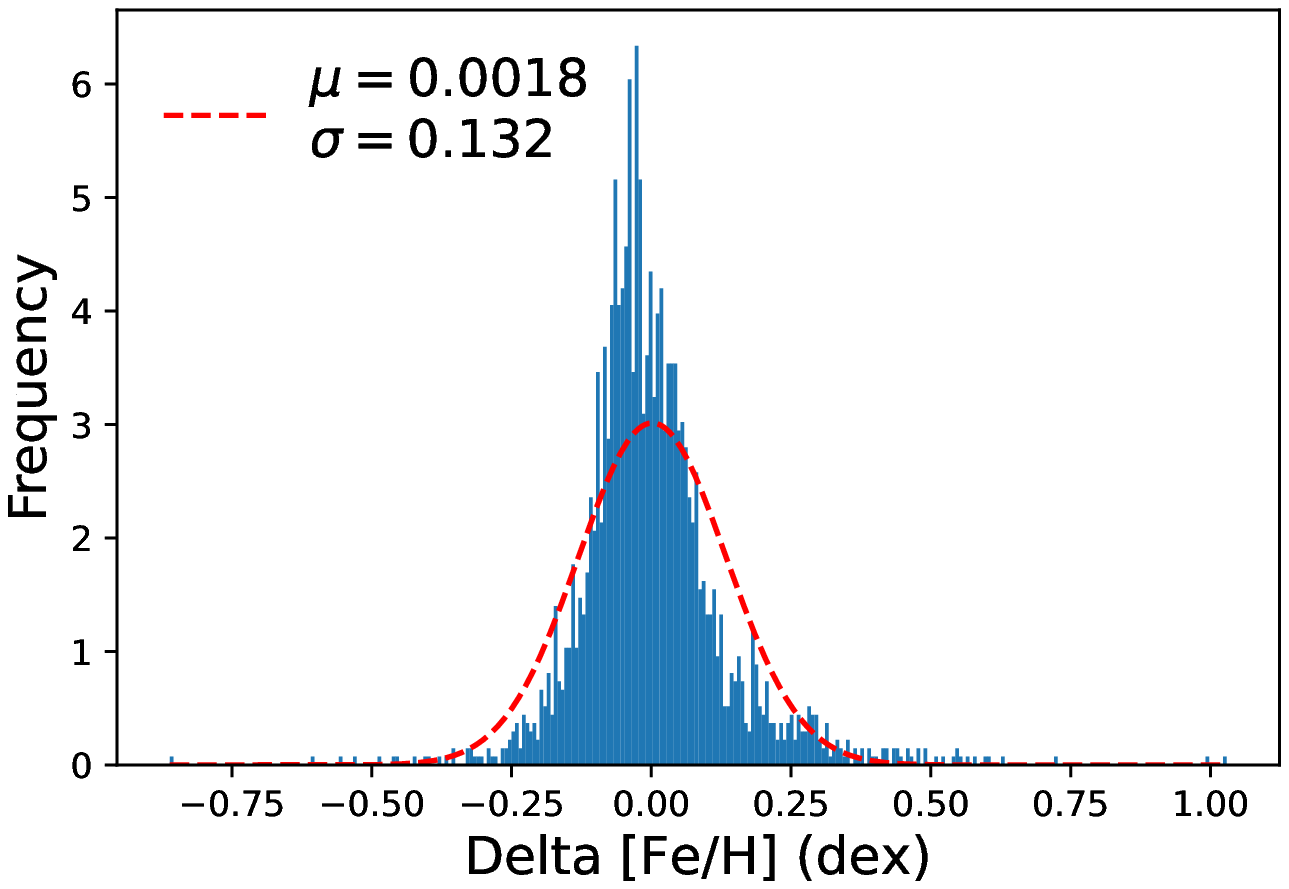}\\
\caption{Similar to Fig.~\ref{fig:apogee_mlp}, the estimation values of LASP for $T_{\texttt{eff}}$, $\log~g$ and [Fe/H] are compared with the corresponding reference values provided by APOGEE. The above is the binary correlation diagram. The following figures are the histogram of the errors. The dashed line represents the Gaussian fit to the residual distribution, and the top of the histogram shows the mean ($\mu$) and standard deviation ($\sigma$) of the error.}
\label{fig:apogee_lamost}
\end{figure}

We compared the parameter estimation results of LASSO-MLP with those of LASP from LAMOST low-resolution spectra, ASPCAP from APOGEE high-resolution spectra respectively. More consistency is shown between  LASSO-MLP estimation and the ASPCAP estimation (the 1st and 2nd rows of the Table~\ref{tab:consistency}). The fundamental principle of LASP is to calculate the difference between each observed flux in the selected wavelength range [3850,5500]\AA ~and the corresponding flux of the reference spectra. The characteristics of accumulation result that the matching result of the LASP is prone to be affected by any noise and distortion on all pixels in this wavelength range. However, the LASSO can adaptively evaluate the combined effects from noise and spectral features on parameter estimation, discard ineffective and redundant components. Therefore, the LASSO-MLP model is less susceptible to noise and distortion, performs more accurately. On the other hand, the MLP can reduce the overfitting risk through early stopping and L2 regularization term. Therefore, the LASSO-MLP model have strong robustness and generalization ability (Fig. ~\ref{fig:apogee_lamost}). Furthermore, the experimental results in Fig. ~\ref{fig:apogee_lamost} much less systematic bias from LASSO-MLP than LASP in the case of teff$<$4000 K and [Fe/H]$<$-1 dex.  Fig.~\ref{fig:apogee_mlp} show some comparison results between LASSO-MLP and ASPACP. The experimental results on various different parameter intervals do not show any obvious bias trend. Therefore, the LASSO-MLP is robust in estimating the stellar atmospheric parameters from the low-SNR spectra.

\begin{table}
\renewcommand\arraystretch{1.5}
\caption{Comparison of LASP and LASSO-MLP with APOGEE.}
\label{tab:consistency}
\centering
\begin{tabular}{|c|c|c|c|c|c|c|c|c|c|}
\hline
 & \multicolumn {3}{c|}  {$T_{\texttt{eff}}$} & \multicolumn {3}{c|}{$\log~g$} & \multicolumn {3}{c|}{[Fe/H]}\\
\cline{2-10}
& MAE (K) & $\mu$ & $\sigma$ & MAE (dex) & $\mu$ & $\sigma$ & MAE (dex) & $\mu$ & $\sigma$ \\ \hline
LASP & 137.6 & -49.51 & 169.6 & 0.195 & -0.063 & 0.257 & 0.091 & 0.0018 & 0.132 \\ \hline
LASSO-MLP  & 84.32 & 0.205 & 164.8 & 0.137 & -0.00084 & 0.217 & 0.063 & -0.00035 & 0.095 \\ \hline
\end{tabular}
\end{table}

\begin{figure}
 \begin{minipage}{7cm}
  \centering
 \includegraphics[width=7cm]{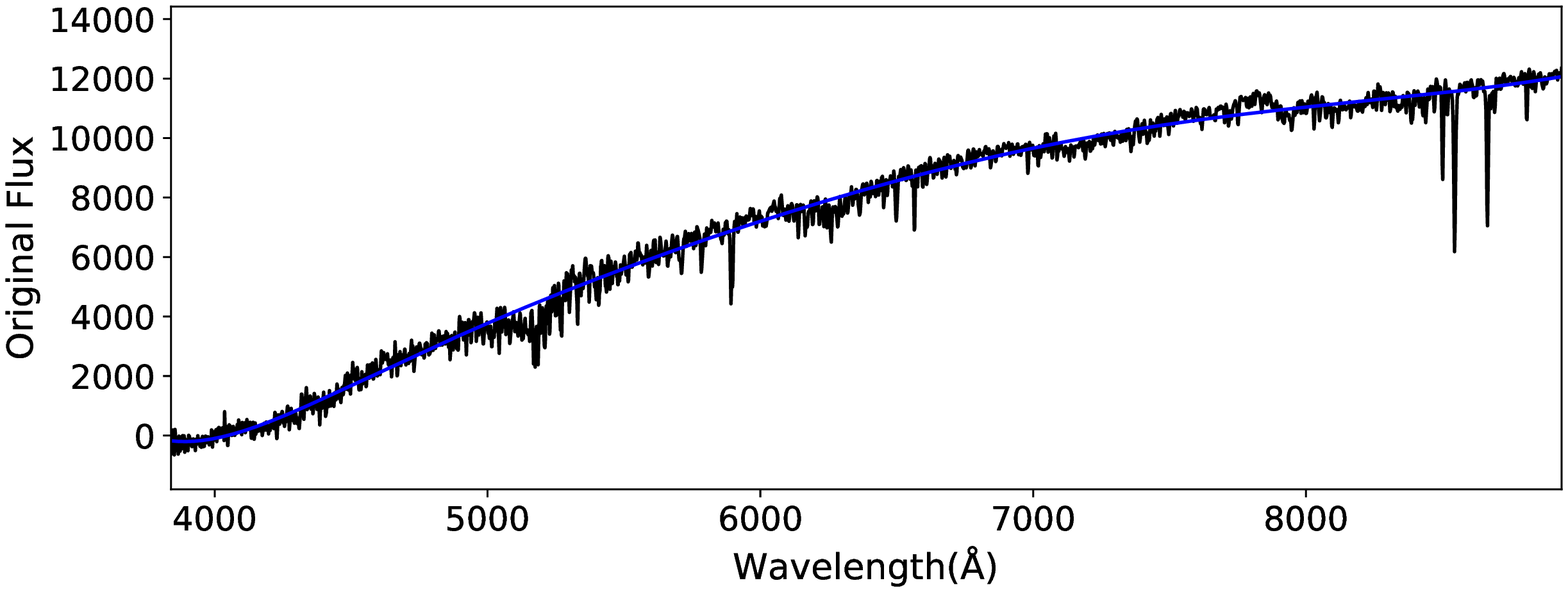}
 \caption*{\quad\quad\quad \small (a) $T_{\texttt{eff}}$ (4062.39 K, 7663.64 K)}
\label{fig:teff_max}
    \end{minipage}
\quad
\begin{minipage}{7cm}
 \includegraphics[width=7cm]{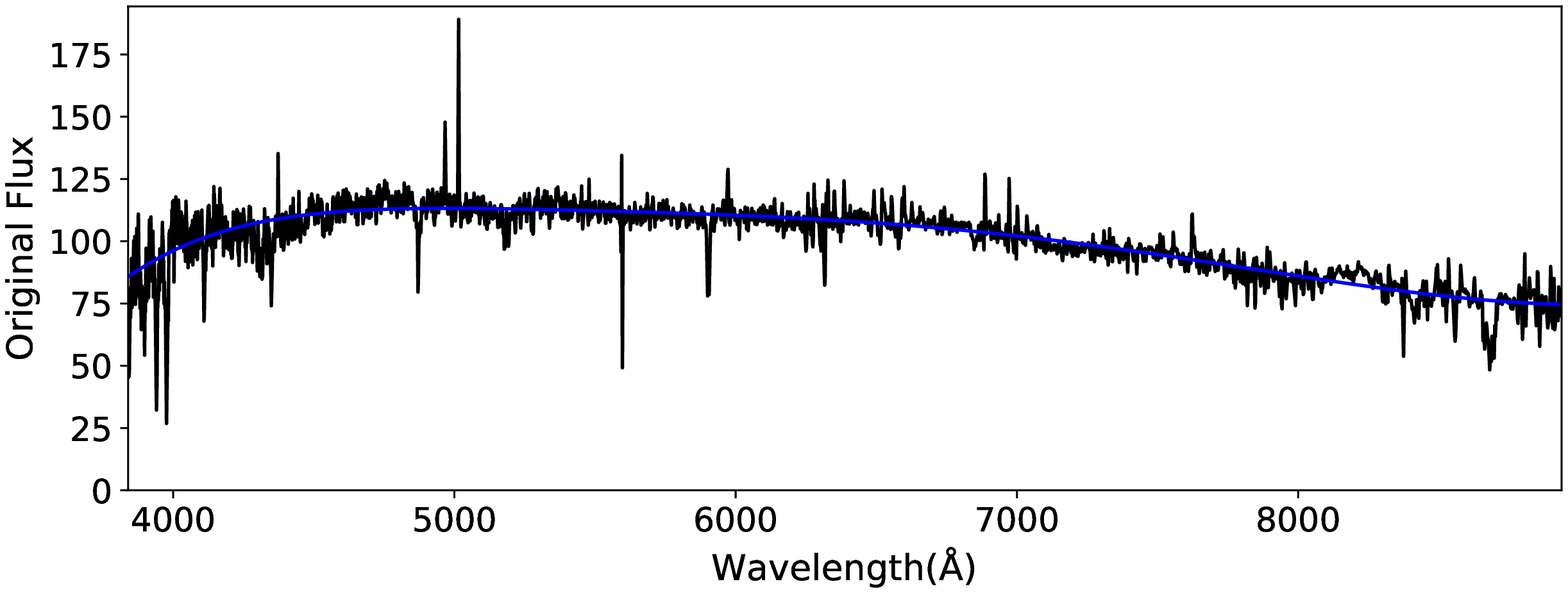}
  \caption*{\quad\quad\quad \small (b) $T_{\texttt{eff}}$ (7581.99 K, 5110.09 K)}
    \label{fig:teff_min}
    \end{minipage}
\\
\begin{minipage}{7cm}
 \includegraphics[width=7cm]{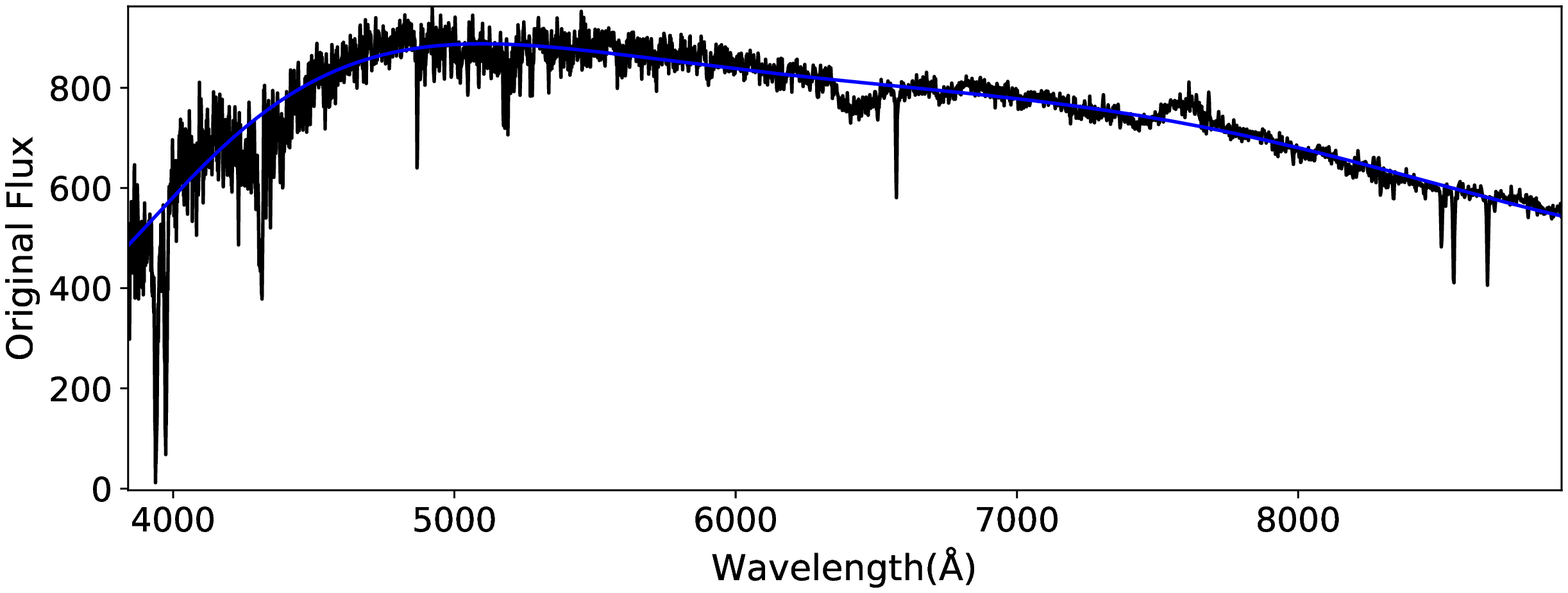}
 \caption*{ \quad\quad\quad \small (c) $\log~g$ms2021-0352fig6c
(2.464 dex, 4.02 dex)}
        \label{fig:logg_max}
    \end{minipage}
\quad
 \begin{minipage}{7cm}
 \includegraphics[width=7cm]{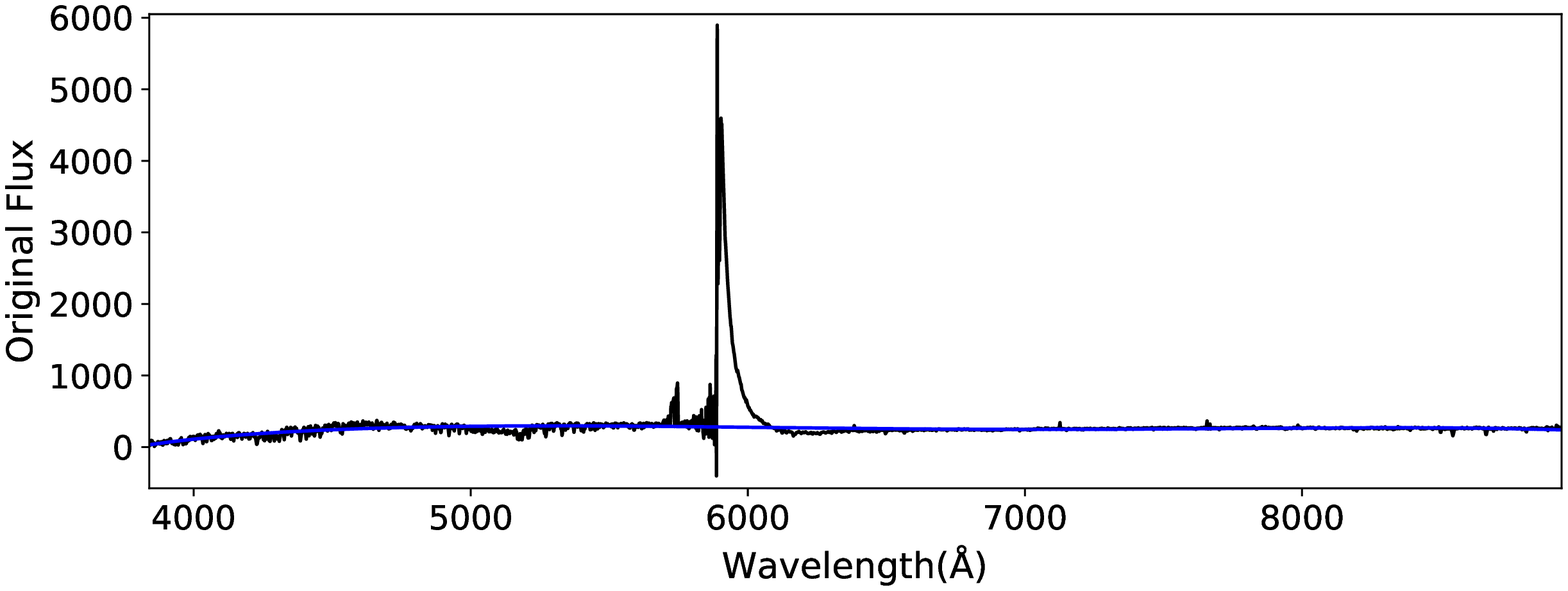}
 \caption*{\quad\quad\quad \small (d) $\log~g$
(4.68 dex, 1.84 dex)}
        \label{fig:logg_min}
    \end{minipage}
\\
 \begin{minipage}{7cm}
 \includegraphics[width=7cm]{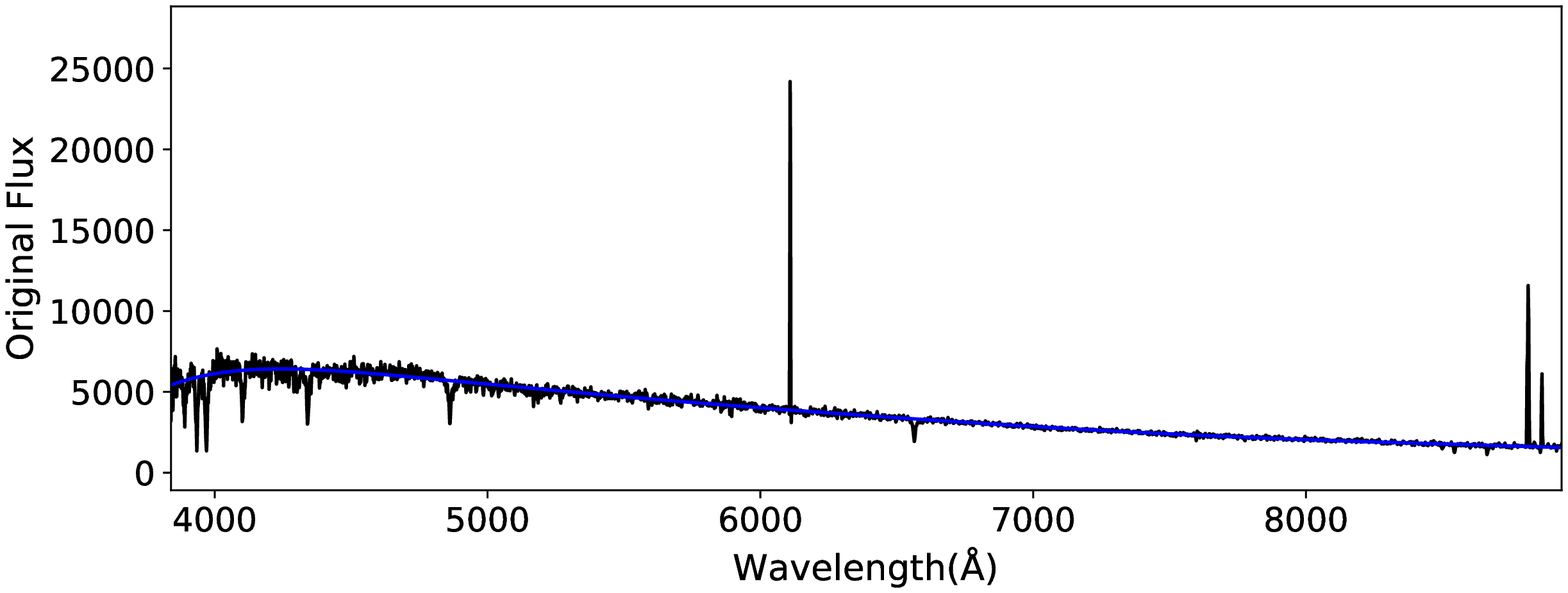}
 \caption*{\quad\quad\quad \small (e) [Fe/H](-0.488 dex,0.052 dex)}
        \label{fig:feh_max}
   \end{minipage}
 \quad
\begin{minipage}{7cm}
 \includegraphics[width=7cm]{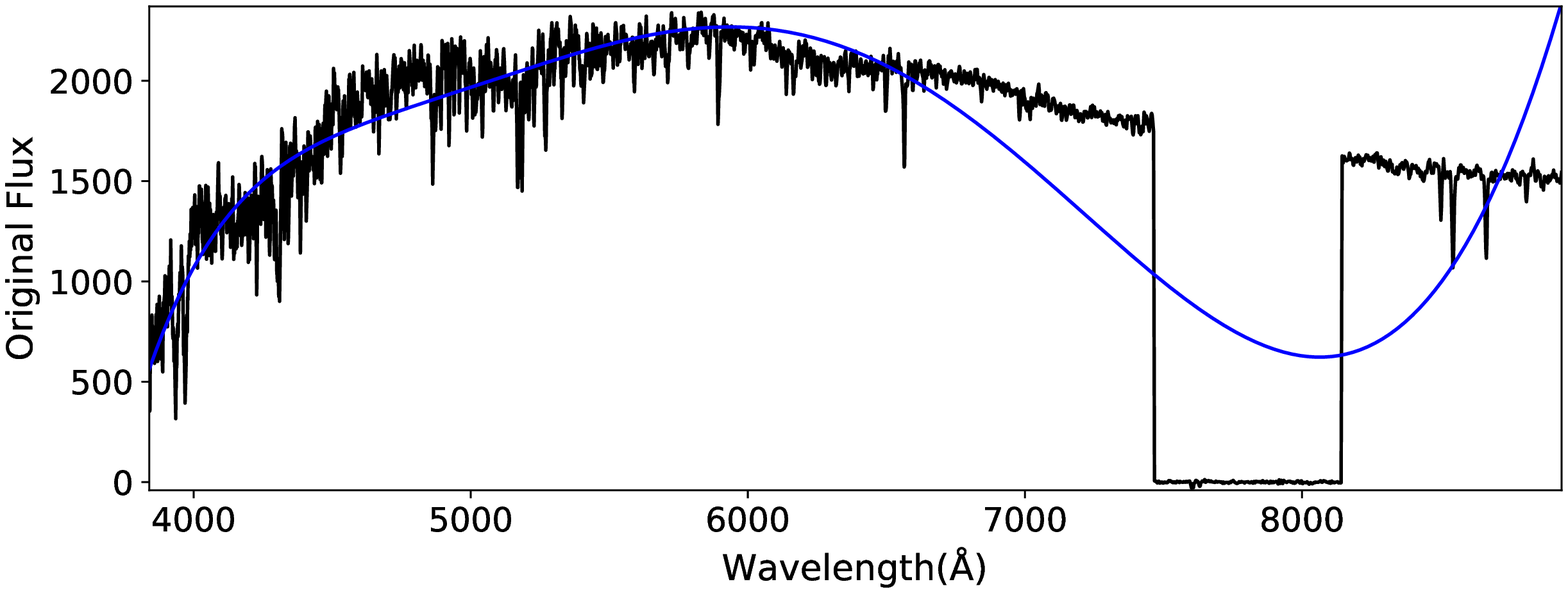}
 \caption*{\quad\quad\quad \small (f) [Fe/H](-0.052 dex, -0.719 dex)}
        \label{fig:feh_min}
   \end{minipage}
 \\
\caption{Six spectra with relatively obvious inconsistencies with the APOGEE catalog. The pair (a, b) in the subfigure caption: a is ASPCAP estimation and b is the MLPNet estimation.}

\label{fig:error}
\end{figure}

However, there are still several spectra with relatively obvious inconsistencies with the APOGEE catalog (Fig.~\ref{fig:apogee_mlp}). These spectra are presented in Fig.~\ref{fig:error}.
Fig.~\ref{fig:error} (a) shows a spectrum with an overestimated $T_{\texttt{eff}}$ by the LASSO-MLP model. The fluxes of the fitted continuum from this spectrum are approximately 0 near 4000 \AA. Therefore, the pre-processing procedure gives some invalid results when dividing the linear interpolation flux $\boldsymbol f$ by the fitted continuum $\boldsymbol f_0$ to normalize the spectrum. That is to say, this spectrum was not properly calibrated during pre-processing. These inappropriate calibrations in preprocessing result in a large deviation in its estimation by the LASSO-MLP model. Fig.~\ref{fig:error} (b) and (c) present two spectra with underestimated $T_{\texttt{eff}}$ and overestimated $\log~g$. These results are due to the large residuals in the sky light emission lines. Fig.~\ref{fig:error} (d) shows a spectrum with an underestimated $\log~g$, and Fig.~\ref{fig:error} (e) presents a spectrum with an overestimated [Fe/H]. These two spectra are affected by some cosmic ray interference. The two cases in Fig.~\ref{fig:error} (d) and (e) indicate that it is necessary to design some mothods detecting the existences of cosmic ray interference and removing/masking them. Fig.~\ref{fig:error} (f) presents a spectrum with an underestimated [Fe/H]. it shows that there is a lot of missing information on the [7500, 8200] \AA  of the spectrum. Therefore, there exist some obvious deviations in their estimations from the LASSO-MLP model.

\section{Improvement on the [Fe/H] estimation for metal-poor stars}\label{sec:metal-poor}

Stars that present lower metallicity than that of the sun, e.g., with [Fe/H]$<-1.0$ are referred to metal-poor stars. They preserve chemical relics of early generations of stars, and thus are important for studying the early formation history of the Milky Way and the universe. However, due to limited survey volume and the near-infrared wavelength coverage, the APOGEE is not able to provide a preferable database for metal-poor stars, and could not cover the stellar parameter space when it comes to [Fe/H]$<-1.5$.
Nevertheless, due to the [Fe/H] coverage range of the common objects between LAMOST low-resolution spectra with 20$\leq$SNR$<$30 and the APOGEE spectra, is [-1.448, 0.429] dex, the accuracy of the trained LASSO-MLP model is not very good in case of [Fe/H]$<$-1.448 dex. Therefore, it is important to find a proper catalogue for low-metallicity stars and design another specific model accordingly.
Fortunately, \citet{2018ApJS..238...16L} has provided the largest catalogue for over 10,000 metal-poor stars based on LAMOST data, and for about 400 of these objects, high-resolution follow-up observations has been performed using the Subaru Telescope, resulting in the largest uniform high-precision database for metal-poor stars (Li et al., under review). Based on this LAMOST/Subaru sample, a catalogue containing 661 LAMOST spectra has been used to establish our new model for metal-poor stars.

To improve the generalization ability of the [Fe/H] estimation model from metal-poor stellar spectrum, a novel reference set is established and denoted by reference set 2. The reference set 2 contains not only all of the 661 metal-poor stellar spectra, but also 600 spectra with [Fe/H]$>$-1.448 dex. These spectra with [Fe/H]$>$-1.448 dex are randomly selected from the LAMOST spectra from the common stars between LAMOST and APOGEE. This reference set is very small. If it is furtherly divided into a training set and a test set, there will be too little data for learning and testing, resulting in a model with poor estimation performance. A small test data set can result in an evaluation result with little statistical significance. Therefore, we designed a five-fold cross validation scheme to build and test the model. In cross validation, we divided the reference set into five mutually exclusive subsets with equal number of spectra, and established five LASSO-MLP models using them. Each model is trained on the reference spectra from four subsets and tested on the reference spectra from the remaining subset. For a spectrum from suspect metal-poor star, we give its [Fe/H] estimation by computing the average of the estimated results from the five models. For convenience, this [Fe/H] estimation model for suspect metal-poor star spectrum is referred to as ensemble LASSO-MLP. This work also trained a LASSO-MLP model using the reference set 2, and this model is denoted by LASSO-MLP$^M$.

Since LASP only provided the [Fe/H] estimations for 255 spectra of the 661 metal-poor stellar spectra, we evaluated the performance of the LASSO-MLP$^M$, ensemble LASSO-MLP model from two aspects. First, we treated the metal-poor star catalog as benchmark, and computed the statistical characteristics of the LASP estimations, LASSO-MLP$^M$, and ensemble LASSO-MLP estimations on the 255 metal-poor stellar spectra (Experiment 1). Second, we computed the inconsistency measures between the ensemble LASSO-MLP estimations and the metal-poor catalog, between the LASSO-MLP$^M$ estimations and the metal-poor catalog on the 661 metal-poor star spectra (Experiment 2).

The experiment 1 shows more consistencies between the LASSO-MLP$^M$ estimation and benchmark than the LASP (the 1st and 2nd rows of the Table~\ref{tab:consistency}). However, the standard deviation ($\sigma$) of the error by the LASSO-MLP$^M$ model is larger than that of LASP. This is due to the small number of samples in the training set, and the complexity of the LASSO-MLP$^M$ model. The LASSO-MLP$^M$ is more complex than the LASP model. Therefore, the LASSO-MLP$^M$ model is prone to overfitting in case of a small training set. The performance evaluation results of the ensemble LASSO-MLP are presented in the 3rd and 5th rows of the Table~\ref{tab:Cross-validation}. It is shown that the ensemble LASSO-MLP significantly improves the accuracy and stability of the parameter estimation. Therefore, we re-estimate the [Fe/H] using the ensemble LASSO-MLP model for 222 spectra with LASSO-MLP [Fe/H] estimation smaller than -1.448 dex. The final estimation results show that there are 209 spectra with [Fe/H]$<$-1.5 dex in the LAMOST DR8 stellar spectra with 20$\leq$SNR$<$30.

For all of the 661 metal-poor stellar spectra, both the LASSO-MLP$^M$ model and the ensemble LASSO-MLP model give [Fe/H] estimations. The experimental results in experiment 2 also show that the the ensemble LASSO-MLP estimations are more consistent with metal-poor star catalog  (the 4th and 5th rows of the Table~\ref{tab:Cross-validation}). Therefore, this work propose the ensemble LASSO-MLP model for estimating the [Fe/H] on the metal-poor stellar spectra.

In theory, a parameter estimation model should be trained and tested on independent samples. In this work, however, the reference data of the metal-poor stars are scarce. Therefore, we did not divide the reference set 2 into a separate training set and a test set. As a result, it is probably that there exist some optimism to a certain in evaluation results on LASSO-MLP$^M$ and ensemble LASSO-MLP (Table~\ref{tab:Cross-validation}).

\begin{table}
\caption{Performance evaluations for the models LASP, LASSO-MLP$^M$ and ensemble LASSO-MLP on estimating the [Fe/H] of metal-poor stellar spectra. The LASSO-MLP$^M$ is the LASSO-MLP trained from the reference set 2 (sec. \ref{sec:metal-poor}).The 1st, 2nd, and 3rd rows shows the performance evaluations of LASP, LASSO-MLP$^M$, and the ensemble LASSO-MLP model on 255 metal-poor stellar spectra with [Fe/H] estimations from LAMOST. The 4th and 5th rows presents the performance evaluations of LASSO-MLP$^M$ and the ensemble LASSO-MLP model on 661 spectra from the metal-poor catalog from Li Haining's team.}
\label{tab:Cross-validation}
\centering
\begin{tabular}{|c|c|c|c|}
\hline
                             & MAE & $\mu$ & $\sigma$ \\ \hline
LASP                         & 0.274 & 0.270 & 0.161 \\ \hline
LASSO-MLP$^M$                    & 0.166 & -0.052 & 0.223 \\ \hline
Ensemble LASSO-MLP           & 0.068 & -0.014& 0.092\\ \hline
LASSO-MLP$^M$*                   & 0.217 & 0.023 & 0.311 \\ \hline
Ensemble LASSO-MLP*          & 0.076&0.0050 & 0.107 \\ \hline
\end{tabular}
\end{table}

\section{conclusion}\label{sec:future}
The proposed models achieve good results in the estimating the stellar atmospheric parameters from LAMOST low resolution spectra with $20\leq$ SNR $<30$. However, there are some limitations to be dealt with in furture. For example, the parameters coverage of the reference spectra is very small. In future, we should try to expand the parameters coverage of the training set.

In this paper, we estimated the stellar atmospheric parameters from 1,162,760 LAMOST low-resolution spectra $20\leq$ SNR $<30$ (LAMOST DR8), and  released it. We also released the model code, trained models, the training spectra and test spectra for reference.

The released catalog is organized in a csv file. This file describes the LASP estimations and the proposed model estimates for all 1,162,760 spectra from LAMOST DR8 with $20\leq$ SNR $<30$. Among them,
 $T_{\texttt{eff}}$$\_$LASP, $\log~g$$\_$LASP, and [Fe/H]$\_$LASP represent stellar atmospheric parameters provided by LASP. $T_{\texttt{eff}}$$\_$MLP, $\log~g$$\_$MLP, [Fe/H]$\_$MLP· represent stellar atmospheric parameters estimated by the proposed scheme.
LAMOST$\_$obsid represents the obsid corresponding to the spectrum. The estimation catalog, learned model, experimental
code, trained model, training data and test data are released on the following websit for scientific exploration and algorithm study: https://github.com/xrli/LASSO-MLP.

\begin{acknowledgements}
The authors thank the reviewer and editor for their instructive
comments. This work was supported by the National Natural Science Foundation of China (Grant Nos. 11973022, 11973049, and U1811464), the Natural Science Foundation of Guangdong Province (No. 2020A1515010710), and the
Youth Innovation Promotion Association of the CAS (id. Y202017).
\end{acknowledgements}

\appendix

\label{lastpage}


\begin{thebibliography}{99}
\bibitem[Jofr{\'e} et al.(2010)]{2010A&A...517A..57J} Jofr{\'e}, P., Panter, B., Hansen, C.~J., et al.\ 2010, \aap, 517, A57

\bibitem[Re Fiorentin et al.(2007)]{2007A&A...467.1373R} Re Fiorentin, P., Bailer-Jones, C.~A.~L., Lee, Y.~S., et al.\ 2007, \aap, 467, 1373

\bibitem[Li et al.(2014)]{2014ApJ...790..105L} Li, X., Wu, Q.~M.~J., Luo, A., et al.\ 2014, \apj, 790, 105

\bibitem[Recio-Blanco et al.(2006)]{2006MNRAS.370..141R} Recio-Blanco, A., Bijaoui, A., \& de Laverny, P.\ 2006, \mnras, 370, 141

\bibitem[Bu \& Pan(2015)]{2015MNRAS.447..256B} Bu, Y. \& Pan, J.\ 2015, \mnras, 447, 256

\bibitem[Xiang et al.(2017)]{2017MNRAS.464.3657X} Xiang, M.-S., Liu, X.-W., Shi, J.-R., et al.\ 2017, \mnras, 464, 3657

\bibitem[Li et al.(2015)]{2015ApJS..218....3L} Li, X., Lu, Y., Comte, G., et al.\ 2015, \apjs, 218, 3

\bibitem[Bailer-Jones(2000)]{2000A&A...357..197B} Bailer-Jones, C.~A.~L.\ 2000, \aap, 357, 197

\bibitem[Katz et al.(1998)]{1998A&A...338..151K} Katz, D., Soubiran, C., Cayrel, R., et al.\ 1998, \aap, 338, 151


\bibitem[Zhang et al.(2019)]{2019PASP..131i4202Z} Zhang, X., Zhao, G., Yang, C.~Q., et al.\ 2019, \pasp, 131, 094202

\bibitem[Koleva et al.(2009)]{2009A&A...501.1269K} Koleva, M., Prugniel, P., Bouchard, A., et al.\ 2009, \aap, 501, 1269

\bibitem[Manteiga et al.(2010)]{2010PASP..122..608M} Manteiga, M., Ord{\'o}{\~n}ez, D., Dafonte, C., et al.\ 2010, \pasp, 122, 608


\bibitem[Ting et al.(2019)]{2019ApJ...879...69T} Ting, Y.-S., Conroy, C., Rix, H.-W., et al.\ 2019, \apj, 879, 69

\bibitem[Lee et al.(2008)]{2008AJ....136.2022L} Lee, Y.~S., Beers, T.~C., Sivarani, T., et al.\ 2008, \aj, 136, 2022

\bibitem[Ness et al.(2015)]{2015ApJ...808...16N} Ness, M., Hogg, D.~W., Rix, H.-W., et al.\ 2015, \apj, 808, 16

\bibitem[Prugniel \& Soubiran(2001)]{2001A&A...369.1048P} Prugniel, P. \& Soubiran, C.\ 2001, \aap, 369, 1048

\bibitem[Li et al.(2017)]{2017RAA....17...36L} Li, X.-R., Pan, R.-Y., \& Duan, F.-Q.\ 2017, Research in Astronomy and Astrophysics, 17, 036

\bibitem[S{\'a}nchez-Bl{\'a}zquez et al.(2006)]{2006MNRAS.371..703S} S{\'a}nchez-Bl{\'a}zquez, P., Peletier, R.~F., Jim{\'e}nez-Vicente, J., et al.\ 2006, \mnras, 371, 703

\bibitem[Boeche et al.(2018)]{2018AJ....155..181B} Boeche, C., Smith, M.~C., Grebel, E.~K., et al.\ 2018, \aj, 155, 181

\bibitem[Zhang et al.(2020)]{2020ApJS..246....9Z} Zhang, B., Liu, C., \& Deng, L.-C.\ 2020, \apjs, 246, 9

\bibitem[Wu et al.(2011)]{2011RAA....11..924W} Wu, Y., Luo, A.-L., Li, H.-N., et al.\ 2011, Research in Astronomy and Astrophysics, 11, 924

\bibitem[Garc{\'\i}a P{\'e}rez et al.(2016)]{2016AJ....151..144G} Garc{\'\i}a P{\'e}rez, A.~E., Allende Prieto, C., Holtzman, J.~A., et al.\ 2016, \aj, 151, 144

\bibitem[Liu et al.(2014)]{2014RAA....14..423L} Liu, C.-X., Zhang, P.-A., \& Lu, Y.\ 2014, Research in Astronomy and Astrophysics, 14, 423-432

\bibitem[Ho et al.(2017)]{2017ApJ...836....5H} Ho, A.~Y.~Q., Ness, M.~K., Hogg, D.~W., et al.\ 2017, \apj, 836, 5

\bibitem[Yang \& Li(2015)]{2015MNRAS.452..158Y} Yang, T. \& Li, X.\ 2015, \mnras, 452, 158

\bibitem[Chen et al.(2015)]{2015RAA....15.1125C} Chen, Y.-Q., Zhao, G., Liu, C., et al.\ 2015, Research in Astronomy and Astrophysics, 15, 1125

\bibitem[Lee et al.(2015)]{2015AJ....150..187L} Lee, Y.~S., Beers, T.~C., Carlin, J.~L., et al.\ 2015, \aj, 150, 187

\bibitem[Xiang et al.(2019)]{2019ApJS..245...34X} Xiang, M., Ting, Y.-S., Rix, H.-W., et al.\ 2019, \apjs, 245, 34

\bibitem[Zhao et al.(2012)]{2012RAA....12..723Z} Zhao, G., Zhao, Y.-H., Chu, Y.-Q., et al.\ 2012, Research in Astronomy and Astrophysics, 12, 723

\bibitem[Wu et al.(2014)]{2014IAUS..306..340W} Wu, Y., Du, B., Luo, A., et al.\ 2014, Statistical Challenges in 21st Century Cosmology, 306, 340

\bibitem[Taylor(2005)]{2005ASPC..347...29T} Taylor, M.~B.\ 2005, Astronomical Data Analysis Software and Systems XIV, 347, 29

\bibitem[Wang et al.(2020)]{2020ApJ...891...23W} Wang, R., Luo, A.-L., Chen, J.-J., et al.\ 2020, \apj, 891, 23]

\bibitem[Luo et al.(2015)]{2015RAA....15.1095L} Luo, A.-L., Zhao, Y.-H., Zhao, G., et al.\ 2015, Research in Astronomy and Astrophysics, 15, 1095

\bibitem[Jofr\'{e} et al.(2019)]{Journal:Paula:2019ARAA}Paula Jofr\'{e}, Ulrike Heiter, and Caroline
Soubiran. Accuracy and precision of industrial stellar abundances. Annual Review of Astronomy and Astrophysics, 57:571-616, 2019

\bibitem[Ren et al.(2016)]{2016RAA....16...45R} Ren, J.-J., Liu, X.-W., Xiang, M.-S., et al.\ 2016, Research in Astronomy and Astrophysics, 16, 45

\bibitem[Ye \& Xie(2010)]{2010arXiv1006.5086Y} Ye, G.-B. \& Xie, X.\ 2010, arXiv:1006.5086

\bibitem[Efron et al.(2004)]{2004math......6456E} Efron, B., Hastie, T., Johnstone, I., et al.\ 2004, math/0406456

\bibitem[Gao \& Li(2017)]{2017ChA&A..41..331G} Gao, W. \& Li, X.-. ru .\ 2017, \caa, 41, 331

\bibitem[Li et al.(2018)]{2018ApJS..238...16L} Li, H., Tan, K., \& Zhao, G.\ 2018, \apjs, 238, 16. doi:10.3847/1538-4365/aada4a
\end{thebibliography}
\end{document}